\documentclass[apj]{emulateapj}
\usepackage{epsfig}
\usepackage{changebar}
\usepackage{natbib}

\bibpunct{(}{)}{;}{a}{}{,}

\shorttitle{DOG Luminosities and Dust Properties}
\shortauthors{Bussmann et al.}

\usepackage{amsmath}
\usepackage{amssymb}

\begin{document}

\title{Infrared Luminosities and Dust Properties of $z \approx 2$ Dust-Obscured
Galaxies}

\author{R. S. Bussmann\altaffilmark{1}, Arjun Dey\altaffilmark{2}, C.
Borys\altaffilmark{3}, V.  Desai\altaffilmark{4}, B. T.
Jannuzi\altaffilmark{2}, E. Le Floc'h\altaffilmark{5}, J.
Melbourne\altaffilmark{6}, K. Sheth\altaffilmark{4}, B. T.  Soifer\altaffilmark{4,6}}

\altaffiltext{1}{Steward Observatory, Department of Astronomy, University of
Arizona, 933 N. Cherry Ave., Tucson, AZ 85721; rsbussmann@as.arizona.edu}
\altaffiltext{2}{National Optical Astronomy Observatory, 950 N. Cherry Ave., Tucson, AZ 85719}
\altaffiltext{3}{Herschel Science Center, California Institute of Technology,
Pasadena, CA 91125}
\altaffiltext{4}{Spitzer Science Center, California Institute of Technology, MS
220-6, Pasadena, CA 91125}
\altaffiltext{5}{Spitzer Fellow, Institute for Astronomy, University of Hawaii, Honolulu, HI 96822}
\altaffiltext{6}{Division of Physics, Math and Astronomy, California Institute 
of Technology, Pasadena, CA 91125}

\newpage

\begin{abstract}

We present SHARC-II 350$\mu$m imaging of twelve 24$\mu$m-bright ($F_{\rm 24\mu
m} > 0.8 \: $mJy) Dust-Obscured Galaxies (DOGs) and CARMA 1mm imaging of a
subset of 2 DOGs.  These objects are selected from the Bo\"otes field of the
NOAO Deep Wide-Field Survey.  Detections of 4 DOGs at 350$\mu$m imply infrared
(IR) luminosities which are consistent to within a factor of 2 of expectations
based on a warm dust spectral energy distribution (SED) scaled to the observed
24$\mu$m flux density.  The 350$\mu$m upper limits for the 8 non-detected DOGs
are consistent with both Mrk~231 and M82 (warm dust SEDs), but exclude cold
dust (Arp~220) SEDs.  The two DOGs targeted at 1mm were not detected in our
CARMA observations, placing strong constraints on the dust temperature: $T_{\rm
dust} > 35-60$~K.  Assuming these dust properties apply to the entire sample,
we find dust masses of $\approx3\times10^{8} \: M_\sun$.  In comparison to
other dusty $z \sim 2$ galaxy populations such as sub-millimeter galaxies
(SMGs) and other {\it Spitzer}-selected high-redshift sources, this sample of
DOGs has higher IR luminosities ($2\times10^{13} \: L_\sun$ vs.
$6\times10^{12} \: L_\sun$ for the other galaxy populations) that are driven by
warmer dust temperatures ($>$35-60~K vs.  $\sim$30~K) and lower inferred dust
masses ($3\times10^{8} \: M_\sun$ vs.  $3\times10^{9} \: M_\sun$).  Wide-field
{\it Herschel} and SCUBA-2 surveys should be able to detect hundreds of these
power-law dominated DOGs.  We use existing {\it Hubble Space Telescope} and
{\it Spitzer}/IRAC data to estimate stellar masses of these sources and find
that the stellar to gas mass ratio may be higher in our 24$\mu$m-bright sample
of DOGs than in SMGs and other {\it Spitzer}-selected sources.  Although much
larger sample sizes are needed to provide a definitive conclusion, the data are
consistent with an evolutionary trend in which the formation of massive
galaxies at $z \sim 2$ involves a sub-millimeter bright, cold-dust and
star-formation dominated phase followed by a 24$\mu$m-bright, warm-dust and
AGN-dominated phase.  

\end{abstract}


\newpage

\section{Introduction} \label{sec:intro} 

In the local universe, the most bolometrically luminous galaxies are dominated
by thermal emission from dust which absorbs ultra-violet (UV) and optical light
and re-radiates it in the infrared (IR) \citep{1986ApJ...303L..41S}.  While
rare locally, these ultra-luminous IR galaxies (ULIRGs) are more common at high
redshift
\citep[e.g.,][]{2001A&A...378....1F,2005ApJ...632..169L,2005ApJ...630...82P}.
Studies combining the improved sensitivity in the IR of the {\it Spitzer Space
Telescope} with wide-field ground-based optical imaging have identified a
subset of this $z \sim 2$ ULIRG population that is IR-bright but also optically
faint
\citep{2004ApJS..154...60Y,2005ApJ...622L.105H,2006ApJ...651..101W,2008ApJ...672...94F,2008ApJ...677..943D}.
In particular, \citet{2008ApJ...677..943D} and \citet{2008ApJ...672...94F}
present a simple and economical method for selecting these systems based on
$R$-band and 24$\mu$m Multiband Imaging Photometer for Spitzer
\citep[MIPS;][]{2004ApJS..154...25R} data.  \citet{2008ApJ...677..943D} employ
a color cut of $R - [24] > 14$ (Vega magnitudes; $\approx$$\, F_\nu (24\mu {\rm
m})/F_\nu (R) > 1000$) to identify objects they call Dust-Obscured Galaxies
(DOGs).  Applied to the 8.6~deg$^2$ Bo\"{o}tes field of the NOAO Deep
Wide-Field Survey (NDWFS) that has uniform MIPS 24$\mu$m coverage for $F_\nu
(24\mu {\rm m}) > 0.3 \:$mJy , this selection yields a sample of $\approx$2600
DOGs, or $\approx$302 deg$^{-2}$.

The extreme red colors and number density of the DOGs imply that they are
undergoing a very luminous, short-lived phase of activity characterized by
vigorous stellar bulge and nuclear black hole growth.  Spectroscopic redshifts
determined for a sub-sample of DOGs using the Deep Imaging Multi-Object
Spectrograph \citep[DEIMOS;][]{2003SPIE.4841.1657F} and the Low Resolution
Imaging Spectrometer \citep[LRIS;][]{1995PASP..107..375O} on the telescopes of
the W.~M.~Keck Observatory (59 DOGs), as well as the Infrared Spectrometer
\citep[IRS;][]{2004ApJS..154...18H} on {\it Spitzer} (47 DOGs) have revealed a
redshift distribution centered on $z \approx 2$ with a dispersion of $\sigma_z
\approx 0.5$ \citep{2008ApJ...677..943D}.  

While DOGs are rare, they are sufficiently luminous ($\approx 90$\% of DOGs
with spectroscopic redshifts have $L_{\rm IR} > 10^{12} \: L_\sun$) that they
may contribute up to one-quarter of the total IR luminosity density from all $z
\sim 2$ galaxies (and over half that from all ULIRGs at $z \sim 2$) and may be
the progenitors of the most luminous ($\sim$4$L^*$) present-day galaxies
\citep{2008ApJ...677..943D,2008ApJ...687L..65B}.  Thus far, the efforts to
estimate the IR luminosities of DOGs have primarily relied upon spectroscopic
redshifts and the observed 24$\mu$m flux density.  \citet{2008ApJ...677..943D}
use an empirical relation between the rest-frame 8$\mu$m luminosity (computed
from the observed 24$\mu$m flux density) and the IR luminosity, derived by
\citet{2007ApJ...660...97C}.  However, there is evidence from sources with
$F_{24\mu{\rm m}} > 0.25 \:$mJy that methods based on only the 24$\mu$m flux
density can overestimate the IR luminosity by factors of 2-10
\citep{2007ApJ...668...45P}.  Results from deep 70$\mu$m and 160$\mu$m imaging
of a sub-sample of 24$\mu$m-bright DOGs are consistent with this, favoring
hot-dust dominated spectral energy distribution (SED) templates like that of
Mrk~231 \citep{2009ApJ...691.1846T} which lead to estimates of the IR
luminosity that are on the low end of the range in $L_{\rm IR}/L_8$ conversion
factors adopted in \citet{2008ApJ...677..943D}.  

In this paper, we present 350$\mu$m and 1mm photometry of a sample of DOGs
whose mid-IR spectral features (silicate absorption, power-law SEDs) suggest
the presence of a strong active galactic nucleus (AGN).  The primary goals of
this study are to measure their IR luminosities and constrain their dust
properties, in particular the dust masses and temperatures.  We also estimate
stellar masses for the sources in the sample using published {\it Hubble Space
Telescope} ({\it HST}) data and {\it Spitzer} InfraRed Array Camera (IRAC)
catalogs from the {\it Spitzer} Deep Wide-Field Survey \citep[SDWFS;
see][]{2009ApJ...701..428A}.  Comparison of the stellar and dust masses
potentially allows us to place constraints on the evolutionary status of these
sources.

In section~\ref{sec:obs}, we present the details of the observations.
Section~\ref{sec:results} presents the DOG SEDs from 0.4$\mu$m to 1mm and IR
luminosity measurements, constraints on the dust temperature, and dust and
stellar mass estimates.  In section~\ref{sec:disc}, we compare our results with
similar studies of sub-millimeter galaxies (SMGs) and {\it Spitzer}-selected
sources from the eXtragalactic First Look Survey (XFLS) and Spitzer Wide
InfraRed Extragalactic (SWIRE) survey.  We present our conclusions in
section~\ref{sec:conc}.

Throughout this paper we assume a cosmology where
$H_0=$70~km~s$^{-1}$~Mpc$^{-1}$, $\Omega_{\rm m} = 0.3$, and $\Omega_\lambda =
0.7$. 

\section{Observations}\label{sec:obs}

\subsection{Sample Selection}\label{sec:sample}

\begin{figure}[!tbp]
\epsscale{1.20}
\plotone{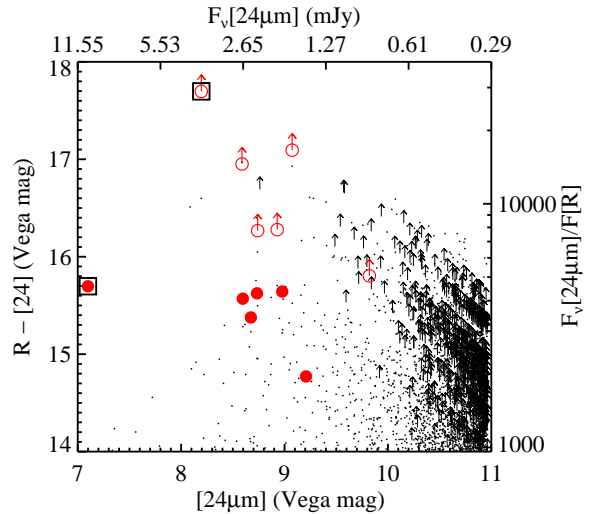}

\caption{ $R - [24]$ color vs. 24$\mu$m magnitude for DOGs in the NDWFS
Bo\"otes field.  Bottom and top abscissae show the 24$\mu$m magnitude and flux
density, respectively.  Left and right ordinates show the color in magnitudes
and the $F_{\rm 24\mu m}/F_{\rm 0.7\mu m}$ flux density ratio, respectively.
Black dots and upward arrows show the full sample of DOGs, with and without an
$R$-band detection, respectively.  The subsample studied in this paper is
represented by red circles (open symbols show sources undetected in the
$R$-band data).  Two sources observed by CARMA at 1mm are highlighted by a
black square.  This plot demonstrates that the sample studied in this paper
probes the 24$\mu$m-bright DOGs over a wide range of $R-[24]$ colors. }

\label{fig:sample}

\end{figure}

\citet{2008ApJ...677..943D} identified 2603 DOGs in the 8.6 deg$^2$ NDWFS
Bo\"otes field, selecting all 24$\mu$m sources satisfying $R - [24] > 14$ (Vega
mag) and $F_{\rm 24\mu m} > 0.3 \:$mJy.  We identified 12 DOGs with
spectroscopic redshifts for follow-up 350$\mu$m imaging (see
Figure~\ref{fig:sample}) with the second-generation Submillimeter High Angular
Resolution Camera (SHARC-II) at the Caltech Sub-millimeter Observatory (CSO).
These targets were selected to have bright 24$\mu$m flux densities ($F_{\rm
24\mu m} \gtrsim 1$~mJy) and a power-law dominated mid-IR SED \citep[based on
{\it Spitzer}/IRAC and 24$\mu$m MIPS photometry; for details, see section 3.1.2
in][]{2008ApJ...677..943D}.  Using the deeper IRAC observations from SDWFS
\citep{2009ApJ...701..428A}, the fraction of DOGs qualifying as power-law
sources ranges from $\approx 10\%$ at $F_{\rm 24\mu m} = 0.3$~mJy to $\approx
60\%$ at $F_{\rm 24\mu m} = 1$~mJy.  As shown in Figure~\ref{fig:sample}, our
sample spans a broad range in $R-[24]$ color ($\approx 14.5 - 17.5$).  {\it
Spitzer}/IRS spectroscopic redshifts have been obtained for these sources based
on the 9.7$\mu$m silicate absorption feature.  Power-law continua and silicate
absorption are typical features of AGN-dominated systems
\citep{2007ApJ...660..167D,2006ApJ...653..101W,2008ApJ...675..960P,2008ApJ...680..119B}.
We note that such systems often exhibit intense star-formation concurrent with
the growth of a super-massive central black hole
\citep[e.g.,][]{2008ApJ...687..848W}.  

Details of our observations are presented in Table~\ref{tab:observations}.  The
effective integration time (column 8 of table~\ref{tab:observations})
represents the time necessary to reach the same noise level given a completely
transparent atmosphere \citep[see][for details]{2008MNRAS.384.1597C}.

We observed two of the twelve DOGs at 1mm using the Combined Array for Research
in Millimeter-wave Astronomy (CARMA) interferometer to search for thermal
emission from cold dust particles.  These were primarily selected to have
robust 350$\mu$m detections to enhance the probability of detection at 1mm and,
in the event of a non-detection, allow useful constraints to be placed on the
dust properties.  The two targets observed with CARMA are
SST24~J142827.2+354127 (S2) and SST24~J143001.9+334538 (S3).
Table~\ref{tab:carmaobservations} presents the date and integration time of the
CARMA observations.

\begin{deluxetable*}{lllllllll}
\tabletypesize{\tiny} 
\tablecolumns{9}
\tablecaption{SHARC-II 350$\mu$m Observations}
\tablehead{
\colhead{} & 
\colhead{} & 
\colhead{RA} & 
\colhead{DEC} & 
\colhead{} & 
\colhead{} &
\colhead{$t_{\rm int}$\tablenotemark{a}} &
\colhead{$t_{\rm eff}$\tablenotemark{b}} &
\colhead{$\sigma_{\rm map}$} \\
\colhead{Source Name} & 
\colhead{ID} & 
\colhead{(J2000)} & 
\colhead{(J2000)} & 
\colhead{$z$} & 
\colhead{UT Year-Month} &
\colhead{(hr)} &
\colhead{(min)} &
\colhead{(mJy)}
}
\startdata
SST24~J142648.9+332927  & S1 &   14:26:48.970 & +33:29:27.56  &  2.00\tablenotemark{c}   & 2006-Apr                      & 1.1 & 4.0 & 20 \\
SST24~J142827.2+354127  & S2 &   14:28:27.190 & +35:41:27.71  &  1.293\tablenotemark{d}  & 2005-Apr/2006-Apr/2007-May  & 3.3 & 11.9& 10 \\
SST24~J143001.9+334538  & S3 &   14:30:01.910 & +33:45:38.54  &  2.46\tablenotemark{e}   & 2005-Apr/2006-Apr/2006-May  & 3.6 & 11.4& 8 \\
SST24~J143025.7+342957  & S4 &   14:30:25.764 & +34:29:57.29  &  2.545\tablenotemark{f}  & 2006-Apr                      & 1.0 & 2.0 & 26 \\
SST24~J143135.2+325456  & S5 &   14:31:35.309 & +32:54:56.84  &  1.48\tablenotemark{c}   & 2007-May                        & 1.2 & 2.0 & 30 \\
SST24~J143325.8+333736  & S6 &   14:33:25.884 & +33:37:36.90  &  1.90\tablenotemark{c}   & 2006-Apr                      & 0.3 & 1.2 & 28 \\
SST24~J143411.0+331733  & S7 &   14:34:10.980 & +33:17:32.70  &  2.656\tablenotemark{g}  & 2005-Jan                    & 3.8 & 12.7& 8 \\
SST24~J143447.7+330230  & S8 &   14:34:47.762 & +33:02:30.46  &  1.78\tablenotemark{e}   & 2007-May                        & 1.8 & 10.5& 10 \\
SST24~J143508.4+334739  & S9 &   14:35:08.518 & +33:47:39.44  &  2.10\tablenotemark{c}   & 2006-Apr                      & 1.0 & 0.5 & 35 \\
SST24~J143539.3+334159  & S10 &  14:35:39.364 & +33:41:59.13  &  2.62\tablenotemark{e}   & 2005-Apr/2006-Apr           & 2.5 & 3.1 & 15 \\
SST24~J143545.1+342831  & S11 &  14:35:45.137 & +34:28:31.42  &  2.50\tablenotemark{c}   & 2006-Apr                      & 0.8 & 3.2 & 22 \\
SST24~J143644.2+350627  & S12 &  14:36:44.269 & +35:06:27.12  &  1.95\tablenotemark{e}   & 2006-Apr/2006-May             & 2.1 & 6.7 & 10 \\
\tablenotetext{a}{Actual on-source integration time.}
\tablenotetext{b}{Effective integration time for a transparent atmosphere \citep{2008MNRAS.384.1597C}.}
\tablenotetext{c}{Redshift from {\it Spitzer}/IRS (Higdon et al. in prep).}
\tablenotetext{d}{Redshift from Keck DEIMOS \citep{2006ApJ...641..133D}.}
\tablenotetext{e}{Redshift from {\it Spitzer}/IRS \citep{2005ApJ...622L.105H}.}
\tablenotetext{f}{Redshift from Keck DEIMOS (Dey et al., in prep.).}
\tablenotetext{g}{Redshift from Keck LRIS \citep{2005ApJ...629..654D}.}
\enddata                                                                          
\label{tab:observations}
\end{deluxetable*}

\begin{deluxetable}{lll}
\tabletypesize{\scriptsize} 
\tablecolumns{3}
\tablecaption{CARMA 1mm Observations}
\tablehead{
\colhead{ID} & 
\colhead{UT Year-Month} &
\colhead{$t_{\rm int}$ (hr)\tablenotemark{a}}
}
\startdata
S2 &   2008-April/May & 10.3 \\
S3 &   2008-April/May & 7.5 \\
\tablenotetext{a}{On-source integration time.}
\enddata                                                                          
\label{tab:carmaobservations}
\end{deluxetable}

\subsection{SHARC-II 350$\mu$m Imaging and Photometry}\label{sec:sharc2}

The SHARC-II observations of the 12 target DOGs were carried out over the
course of five separate observing runs from 2005 January to 2007 May.  Data
were collected only when the atmospheric opacity was low and conditions were
stable ($\tau_{\rm 225 GHz} < 0.06$).  Pointing, focus checks, and calibration
were performed every hour, using ULIRG Arp~220 as a calibrator ($F_{350 \mu
{\rm m}} = 10.2 \pm 1.0 \:$Jy).  Other secondary calibrators (CIT6, CRL2688,
3C345) were used occasionally to verify the Arp~220 calibrations.  A
non-connecting Lissajous pattern was used to modulate the telescope pointing
with amplitudes of 15$\arcsec$-20$\arcsec$ and periods of 10-20~s.  The
observations made use of the CSO Dish Surface Optimization System to optimize
the dish surface accuracy and beam efficiency \citep{2006SPIE.6275E..21L}.

Data were reduced using the Comprehensive Reduction Utility for SHARC-II
(CRUSH) software package with the `deep' option to optimize the signal-to-noise
ratio (S/N) for faint ($<$ 100~mJy) point sources \citep{2006PhDT........28K}.
The output map has a pixel scale of 1$\farcs$62~pixel$^{-1}$, and is smoothed
with a 9$\arcsec$ gaussian beam, resulting in an effective image FWHM of
12$\farcs$4.  

A 20$\arcsec$ diameter aperture was used for photometry to compute the
instrumental flux density of each source.  The sky level and photometric
uncertainty were computed by measuring the mean and RMS in $\approx$10
off-source 20$\arcsec$ diameter apertures.  The same procedure was applied to
the calibration images, and a scaling factor was derived that converts the
instrumental flux density to a physical flux density (using this method, no
subsequent aperture correction is required as long as both the science and
calibration targets are unresolved and measured in the same aperture).

The aperture photometry is consistent with peak flux density measurements in
all but one source, SST24~J142648.9+332927.  This source has a radial profile
that is significantly more extended than the point spread function of the final
map, which results in the peak flux underestimating the aperture flux density
measurement.  The extended structure in the image is more likely to be noise
than signal, so in this case we report the peak flux measurement, which is
formally a non-detection.


Flux boosting of low S/N sources can be an important effect in wide-field
surveys where source positions are not known a priori
\citep[e.g.,][]{2005MNRAS.357.1022C}.  However, because we know our source
positions at the $<1\arcsec$ level (from MIPS and IRAC centroids), flux
boosting is not a significant effect, and so we do not apply any such
corrections to our measurements.  Our approach follows that adopted by
\citet{2006ApJ...643...38L} and \citet{2006ApJ...650..592K} in their 350$\mu$m
follow-up imaging of SMGs.

\subsection{CARMA 1mm Imaging and Photometry}\label{sec:carma}

The CARMA observations were obtained between 2008 April 7 and May 1 in the
C-array configuration (beamsize is $\approx 2 \times 1$ sq. arcsec).  A total
of 7.5 hours and 10.3 hours of integration time in good 1mm weather conditions
were spent on sources SST24~J143001.9+334538 (S3) and SST24~J142827.2+354127
(S2), respectively.  These sources were selected primarily because of their
robust (S/N$> 4.5$) detections at 350$\mu$m.  In addition, source S3 is
detected with {\it Spitzer}/MIPS at 70$\mu$m and 160$\mu$m
\citep{2009ApJ...691.1846T}, while S2 is the subject of a detailed
spectroscopic study \citep{2006ApJ...641..133D}.  

System temperatures were in the range 250-400~K.  A correlator configuration
was used with three adjacent 15$\times$31~MHz bands centered on 220~GHz, the
frequency at which the CARMA 1mm receivers are most sensitive.  The quasar
J1310+323 (chosen for its spatial proximity) was observed every 15 minutes for
amplitude and phase calibration.  Quasars 3C~273 and MWC~349 were used for
pointing, pass-band calibration, and flux calibration.

Data were reduced using the Multichannel Image Reconstruction, Image Analysis,
and Display (MIRIAD) software package \citep{1995ASPC...77..433S}.  Visual
inspection of visibilities as a function of baseline length allowed us to
identify and flag spurious data.  A cleaned map was generated for each track of
integration time (ranging from 1 to 5 hours in length) and these tracks were
coadded together using a weighted mean to obtain a final image of these sources.
No detections were found in either case.  Both sources are unresolved in the
IRAC images, and NICMOS/F160W imaging of S3 indicates a size of $\lesssim
0\farcs5$, implying that it is very unlikely any emission was resolved out by
the interferometer.  The {\sc imstat} routine from MIRIAD was used to determine
the noise level in the co-added images where we expected to see the source.
Table~\ref{tab:photometry} shows the photometry from 24$\mu$m to 1mm, where
available.  Non-detections are given as 3-$\sigma$ upper limits.

\subsection{Optical, near-IR, mid-IR, and far-IR Photometry}\label{optirphot}

The optical and near-IR photometry used in this paper to estimate stellar
masses are based on high spatial-resolution {\it HST} data (WFPC2/F606W,
ACS/F814W, and NIC2/F160W) published in \citet{2009ApJ...693..750B}. The {\it
HST} data allow the separation of an unresolved nuclear component (flux on
scales $\lesssim 1~$kpc) from a more spatially extended component. Because the
AGN contribution in the rest-frame UV to optical is uncertain, the photometry
of the extended component is used here (measured with 2$\arcsec$ diameter
apertures) to ensure that our stellar mass estimates are not biased by the
presence of an obscured AGN \citep[for details on how the extended component is
computed, see][]{2009ApJ...693..750B}. Additionally, 4$\arcsec$ diameter
aperture photometry in the optical ($B_W$, $R$, and $I$) from the NDWFS is
shown in the SEDs of the objects in this sample \citep[details on how the
photometry is computed may be found in][]{2009ApJ...693..750B}.

The mid-IR photometry used in this paper are from the publicly available Data
Release 1.1 (DR1.1) catalogs from the SDWFS IRAC coverage of the Bo\"otes field
(Ashby et al.~2009). The SDWFS catalogs incorporate the earlier IRAC Shallow
Survey of the Bo\"otes field undertaken by the IRAC guaranteed time observation
(GTO) programs \citep{2004ApJS..154...48E}. We identified IRAC counterparts of
the DOGs in this paper from the SDWFS catalogs using a 3$\arcsec$ search radius
centered on the 24$\mu$m position (the MIPS 24$\mu$m 1-$\sigma$ positional
uncertainty is $1\farcs2$). All of the DOGs in this paper have IRAC
counterparts, detected at $>4\sigma$ in all four IRAC channels.  We use the
4$\arcsec$ (rather than the 6$\arcsec$) diameter aperture photometry from the
DR1.1 SDWFS catalog to reduce contamination from nearby sources.  We note that
aperture corrections derived from isolated, bright stars have been applied to
the SDWFS catalogs.

Finally, 24, 70, and 160$\mu$m data over 8.61 deg$^2$ of the Bo\"otes field are
available from GTO programs. The data were reduced by the MIPS GTO team and
reach 1$\sigma$ rms depths of $51~\mu$Jy, 5~mJy, and 18~mJy at 24, 70, and
160$\mu$m, respectively. Details of the GTO surveys, such as mapping strategy,
data reduction, and source catalogs, will be discussed elsewhere. In addition,
several of the DOGs in this paper were targeted for deeper MIPS photometric
observations by {\it Spitzer} General Observer program 20303 (P.I. E.
LeFloc'h), and the results are reported in \citet{2009ApJ...691.1846T}. We use
the \citet{2009ApJ...691.1846T} measurements where they are available.

\begin{deluxetable*}{llllllll}
\tabletypesize{\scriptsize} 
\tablecolumns{8}
\tablecaption{Photometry\tablenotemark{a}}
\tablehead{
\colhead{} & 
\colhead{$R - [24]$} &
\colhead{$F_{24\mu{\rm m}}$} & 
\colhead{$F_{70\mu{\rm m}}$} & 
\colhead{$F_{160\mu{\rm m}}$} & 
\colhead{$F_{350\mu{\rm m}}$} & 
\colhead{$F_{1.2{\rm mm}}$} &
\colhead{$F_{20{\rm cm}}$\tablenotemark{b}} \\
\colhead{Source Name} &
\colhead{(Vega mag)} &
\colhead{(mJy)} &
\colhead{(mJy)} &
\colhead{(mJy)} &
\colhead{(mJy)} &
\colhead{(mJy)} &
\colhead{(mJy)} 
}
\startdata
S1  & $>$16.1  &  2.33$\pm$0.07 & ---  &  ---   & $< 66$  &  --- & --- \\
S2  & 15.7  & 10.55$\pm$0.13 & ---  &  $<$45\tablenotemark{c} & 74$\pm$13 & $<$1.5 & --- \\
S3  & $>$17.4  &  3.84$\pm$0.06 & 9.3$\pm$2.3\tablenotemark{d} & 65$\pm$11\tablenotemark{d} & 41$\pm$13 & $<$1.8 & $0.42\pm0.04$ \\
S4  & 15.1  &  2.47$\pm$0.05 & ---  &  --- & $<$81 & --- & $0.23\pm0.03$  \\
S5  & 14.5  &  1.51$\pm$0.05 & ---  &  --- & $<$100 & --- & $0.54\pm0.12$ \\
S6  & 15.4  &  1.87$\pm$0.06 & ---  &  --- & $<$137 & --- & $0.20\pm0.03$ \\
S7  & 12.4\tablenotemark{e}  &  0.86$\pm$0.05 & $<$25\tablenotemark{f} & $<$90\tablenotemark{f} & 37$\pm$13 & --- & --- \\
S8  & $>$16.8  &  1.71$\pm$0.04 & ---  & ---  & 45$\pm$12 & --- & $0.31\pm0.06$ \\
S9  & 15.3  &  2.65$\pm$0.08 & ---  & ---  & $<$150 & --- & $0.24\pm0.04$ \\
S10 & $>$16.7  &  2.67$\pm$0.06 & $<$8.1\tablenotemark{d} & $<$38\tablenotemark{d} & $<$50 & --- & --- \\
S11 & $>$16.0  &  1.95$\pm$0.05 & ---  & ---  & $<$60 & --- & --- \\
S12 & 15.4  &  2.34$\pm$0.05 & 9.1$\pm$2.5  & 43$\pm$12  & $<$34 & --- & $5.1\pm0.2$ \\
\tablenotetext{a}{Upper limits quoted are 3$\sigma$ values.}
\tablenotetext{b}{Photometry from Westerbork Synthesis Radio Telescope imaging \citet{2002AJ....123.1784D}.}
\tablenotetext{c}{Photometry from \citet{2006ApJ...641..133D}.}
\tablenotetext{d}{Photometry from \citet{2009ApJ...691.1846T}.}
\tablenotetext{e}{$R$-band photometry includes diffuse emission from Ly-$\alpha$ nebula \citep{2005ApJ...629..654D}.}
\tablenotetext{f}{Photometry from \citet{2005ApJ...629..654D}.}
\enddata                                                                          
\label{tab:photometry}
\end{deluxetable*}

\begin{figure*}[!tbp]
\epsscale{1.00}
\plotone{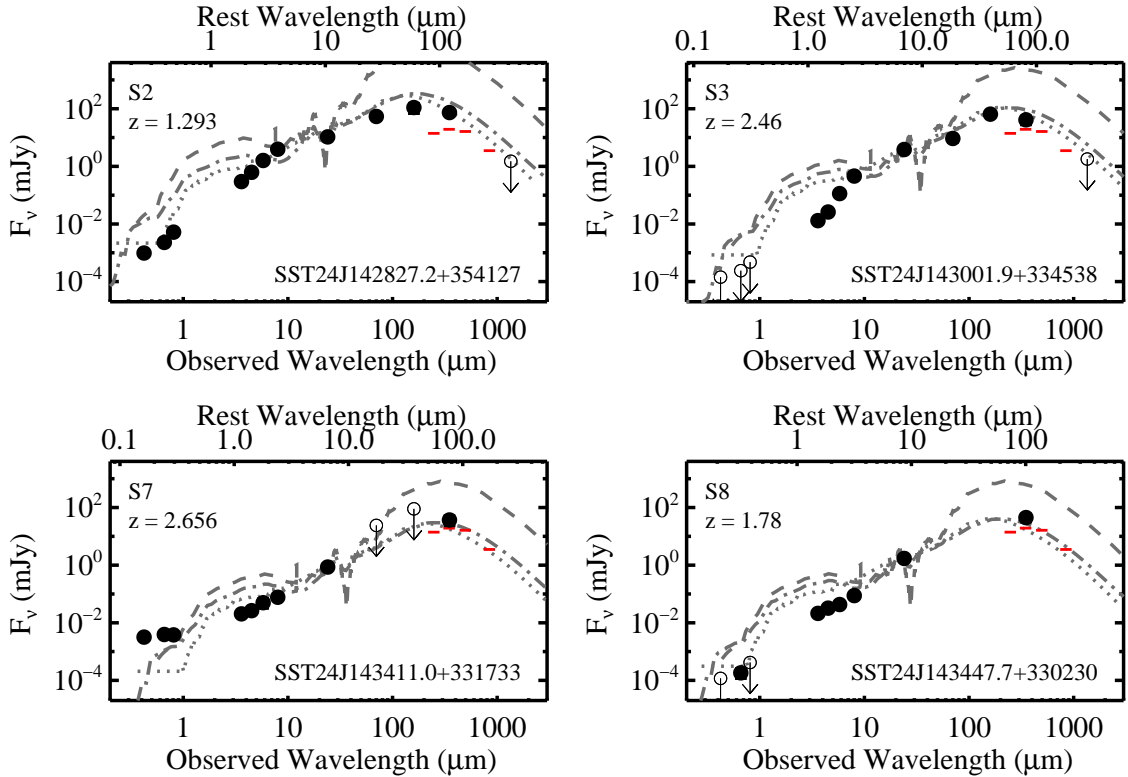}

\caption{ SEDs of 5 DOGs detected by SHARC-II at 350$\mu$m.  Dotted, dashed,
and dot-dashed lines show the Mrk~231, Arp~220, and M82 template SEDs,
respectively, placed at the appropriate redshift and scaled to match the
observed 24$\mu$m flux density.  Red horizontal lines show the 5-$\sigma$
sensitivity limits (ignoring confusion) from the planned wide-field {\it
Herschel} surveys at 250, 350, and 500$\mu$m, and SCUBA-2 surveys at 850$\mu$m.
The cool dust SED of Arp~220 significantly overpredicts the 350$\mu$m flux
density in all cases.  The warm dust SED of M82 provides a better fit in the
far-IR, but Mrk~231 provides the best fit in both the far-IR and the optical.}

\label{fig:seds_detect}

\end{figure*}

\begin{figure*}[!tbp]
\epsscale{1.00}
\plotone{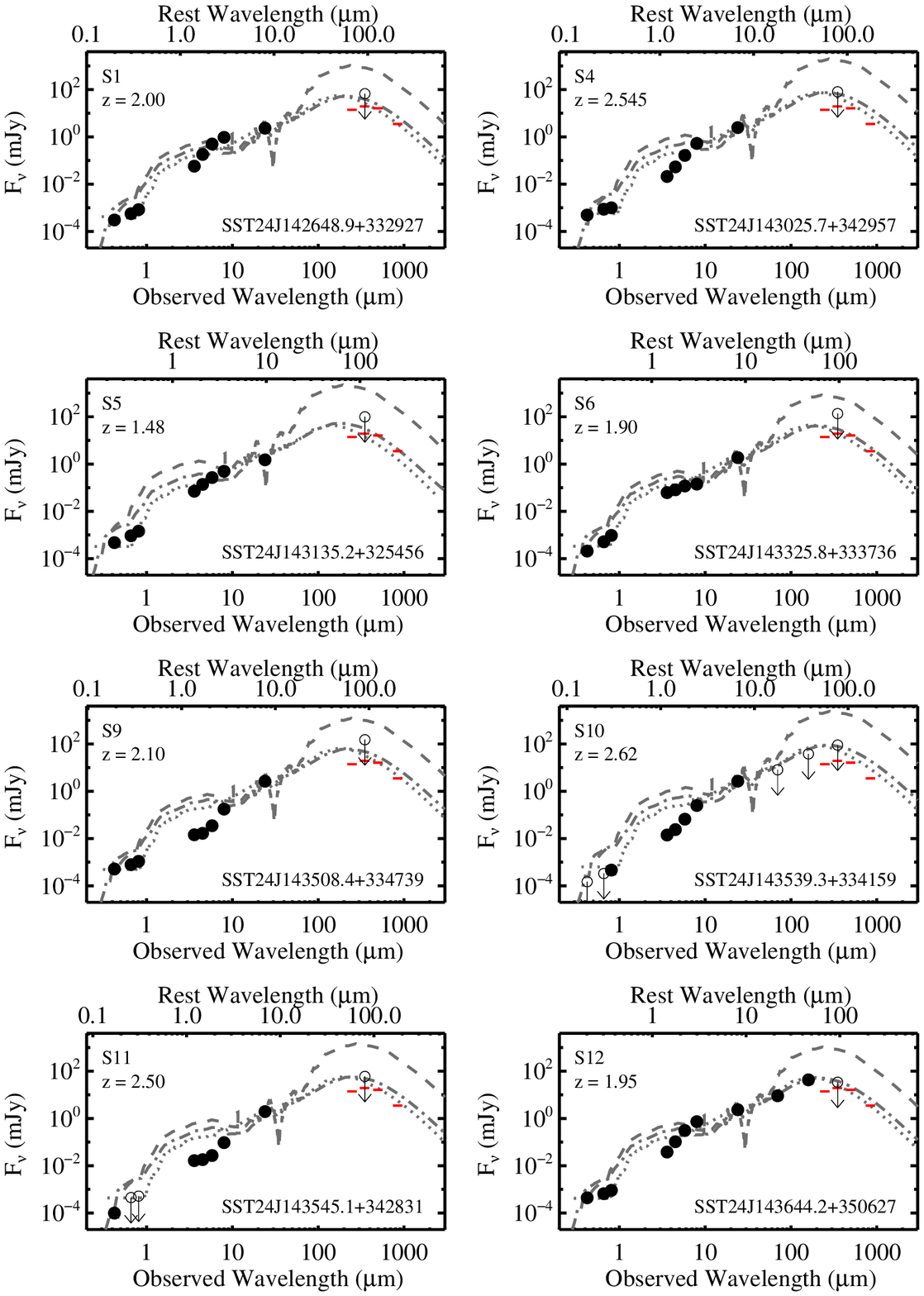}

\caption{ Same as Figure~\ref{fig:seds_detect}, except showing SEDs of 7 DOGs
not detected by SHARC-II at 350$\mu$m.  The limits at 350$\mu$m are all
inconsistent with the Arp~220 SED.  The 250$\mu$m channel of {\it Herschel}
should be very efficient for detecting power-law DOGs in wide-field surveys,
assuming that Mrk~231 is an appropriate representation of the far-IR SED.}

\label{fig:seds_nondetect}

\end{figure*}

\section{Results}\label{sec:results}

In this section, we present SEDs from 0.4$\mu$m to 1mm for each source in our
sample and compare with local starburst (M82) and ULIRG (Arp~220 and Mrk~231)
templates.  Our approach is to artificially redshift the local galaxy templates
and normalize them to match the DOG photometry at observed-frame 24$\mu$m.
This allows a simple, qualitative comparison of DOGs and galaxies with
properties ranging from warm dust, star-formation dominated (M82), to cool
dust, star-formation dominated (Arp~220), to warm dust, AGN-dominated
(Mrk~231).  We will use the SED that provides the best fit over the sampled
wavelength range (observed optical through sub-mm) to estimate the IR
luminosities of the DOGs.

Later in this section, we use our limits at 1mm from CARMA to place constraints
on the dust temperatures and limits on the dust masses of DOGs.  Finally, we
use {\it HST} and {\it Spitzer}/IRAC data to estimate stellar masses of DOGs.

\subsection{Qualitative SED Comparison}\label{sec:seds}

Figures~\ref{fig:seds_detect} and \ref{fig:seds_nondetect} show the SEDs of the
sources, divided respectively into those with and those without detections at
350$\mu$m.  Note that the rest-frame UV photometry for source S7 is
contaminated by emission from nearby sources and will be treated in more detail
in a future paper (Prescott et al., in prep.).  

Overplotted in each panel are M82 \citep{1998ApJ...509..103S}\footnote{We use a
slightly updated SED obtained from
http://adlibitum.oat.ts.astro.it/silva/grasil/modlib/modlib.html}, Mrk~231
(Chary 2008, private communication), and Arp~220 \citep{2008arXiv0810.4150R}
templates, placed at the appropriate redshift and scaled to match the flux
density observed in the MIPS 24$\mu$m band.  The scaling factors derived for
the three templates range over 200-900, and 2-10, 70-700, respectively (the
deep silicate absorption feature in Arp~220 and the strong PAH emission feature
of M82 make the scaling factors closer to each other than a simple estimate
based on the ratio of the IR luminosities would imply).  For the Arp~220 and
Mrk~231 templates, we have interpolated the spectrum in the UV to match Galaxy
Evolution Explorer photometry (in the case of Arp~220) and International
Ultraviolet Explorer data as well as {\it HST} Faint Object Spectrograph data
\citep[in the case of Mrk~231;][]{1987AJ.....93...14H,2002ApJ...569..655G}.

These templates were chosen because they sample a range of dust temperatures
and AGN/starburst contributions.  M82 is one of the closest ($d_L = 3.86
\:$Mpc) galaxies undergoing a starburst, as it was triggered by a recent
interaction with M81.  Although it is less luminous than DOGs \citep[$L_{\rm
IR} \approx 6 \times 10^{10} \: L_\sun$; \footnote{$L_{\rm IR}$ is the
luminosity integrated over 8-1000$\mu$m}][]{2003AJ....126.1607S}, its nucleus
is dominated by a warm dust component \citep[$T_{\rm dust} = 48
\:$K;][]{1994MNRAS.270..641H}.  Arp~220 is a nearby ($d_L =  77.3 \:$Mpc) ULIRG
\citep[$L_{\rm IR} \approx 1.6 \times 10^{12} \:
L_\sun$;][]{2003AJ....126.1607S} dominated by cold dust \citep[$T_{\rm dust} =
35 \:$K;][]{1996MNRAS.278.1049R}.  Mrk~231 is another nearby ($d_L = 175.1
\:$Mpc) ULIRG \citep[$L_{\rm IR} \approx 3.2 \times 10^{12} \:
L_\sun$;][]{2003AJ....126.1607S}, but has a warm dust \citep[$T_{\rm dust} = 51
\:$K;][]{2007ApJ...662..284Y} SED dominated by an obscured AGN.

Qualitatively, the Mrk~231 template provides a much better fit than the Arp~220
template to the 350$\mu$m photometry in every case.  M82 fits the 24$\mu$m and
350$\mu$m photometry reasonably well (although not as well as Mrk~231), but it
fares poorly in the mid-IR and optical, where a strong stellar component in M82
is not seen in the DOGs in this sample (which are dominated by a power-law in
the mid-IR).  Additionally, M82 shows strong PAH emission which is not seen in
the power-law DOGs.

The red horizontal bars in Figures~\ref{fig:seds_detect} and
\ref{fig:seds_nondetect} show 5-$\sigma$ limits (ignoring confusion) from
planned wide-field ($> 8 \:$deg$^2$) surveys with the {\it Herschel} Space
Observatory (shown for the channels at 250, 350, and 500$\mu$m) and with the
Sub-mm Common-User Bolometer Array-2 (SCUBA-2) instrument at 850 $\mu$m.  Most
of the power-law DOGs studied in this paper have SEDs that peak around
observed-frame 250$\mu$m, which is where the {\it Herschel} wide-field maps
will be the deepest (5-$\sigma$ limit of 14~mJy).  If all of the
24$\mu$m-bright DOGs are detected at 250$\mu$m in the two wide-field surveys
that are planned to reach the depths assumed here (Lockman Hole east, 11
deg$^2$; Extended Chandra Deep Field South, 8 deg$^2$), then a total of
$\approx 600$ power-law DOGs should be detected in the 250$\mu$m {\it Herschel}
catalogs of these two fields.  The SCUBA-2 surveys of these fields should be
deep enough (5-$\sigma$ limit of 3.5~mJy at 850$\mu$m) to detect many of these
sources, allowing dust temperature constraints to be placed on a statistically
significant sample of these rare, important objects.

\begin{figure*}[!tbp]
\epsscale{0.80}
\plotone{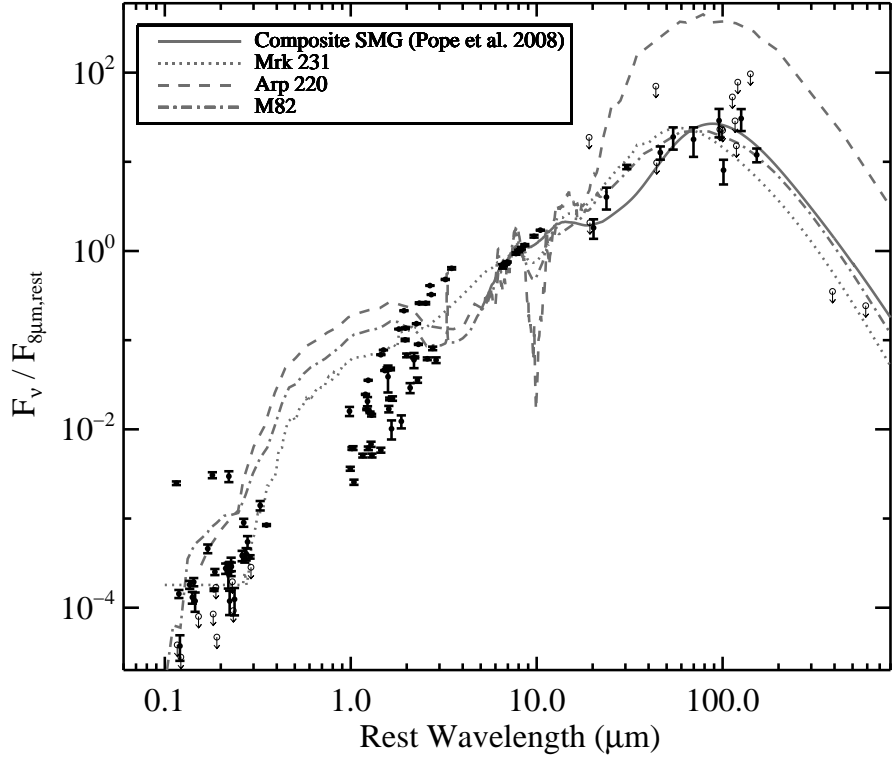}

\caption{ Optical through sub-mm SEDs of DOGs in the SHARC-II sample.  Flux
densities have been normalized by the rest-frame 8$\mu$m flux density, computed
from the observed 24$\mu$m flux density.  Of the three local galaxy templates
shown, Mrk~231 provides the best fit over the rest-frame UV through sub-mm
range because it has a warm dust SED (unlike Arp~220) and because it lacks a
strong stellar component (unlike both Arp~220 and M82).  None of the template
SEDs match the steepness of the rest-frame near-IR photometry of the power-law
DOGs.  This may indicate that obscured AGN dominate stellar emission to a
greater extent in power-law DOGs than in Mrk~231.}

\label{fig:combo_seds}

\end{figure*}

Figure~\ref{fig:combo_seds} shows all of the DOG SEDs on the same plot,
normalized by the rest-frame 8$\mu$m flux density (which is estimated from the
observed-frame 24$\mu$m flux density by assuming a power law of the form $F_\nu
\propto \nu^\alpha$, where $\alpha=-2$).  The SEDs M82, Arp~220, Mrk~231, and
a composite SMG template SED spanning mid-IR to sub-mm wavelengths are also
shown.  The composite SMG template is derived from bright ($F_{850\mu {\rm m}}
> 5 \:$mJy) SMGs from the Great Observatories Origins Deep Survey North
(GOODS-N) field with mid-IR spectra \citep{2008ApJ...675.1171P}.

One striking feature of this plot is the steep slope shown by DOGs in the
rest-frame 1-4$\mu$m.  Whereas Mrk~231, M82, and Arp~220 all exhibit a bump in
the 1-2$\mu$m regime, no such feature is apparent in the DOG SEDs.  This could
be due to the presence of an obscured AGN outshining the stellar light, in the
rest-frame near-IR.  This is in constrast to rest-frame UV and optical
wavelengths, where {\it HST} imaging has revealed that stellar light appears to
dominate \citep{2009ApJ...693..750B}.  As noted previously, DOGs have far-IR to
mid-IR flux density ratios more similar to Mrk 231 than Arp 220.  The composite
SMG template overpredicts the far-IR flux for a given mid-IR flux in all cases
where we have 350$\mu$m detections.  We note that adding an additional warm
dust component ($T_{\rm dust} = 350$~K; possibly powered by an AGN) to the
composite SMG SED \citep[as was done in][]{2008arXiv0808.2816P} improves the
quality of the fit over the rest-frame 8-100$\mu$m.  However, this composite
SMG + AGN template retains a strong cool dust ($T_{\rm dust} \approx 30 \:$K)
component that over-predicts the amount of emission at 1mm.  If this type of
SED was appropriate for the power-law DOGs investigated in this paper, they
would have been easily detected by CARMA.

\begin{figure*}[!tbp]
\epsscale{0.80}
\plotone{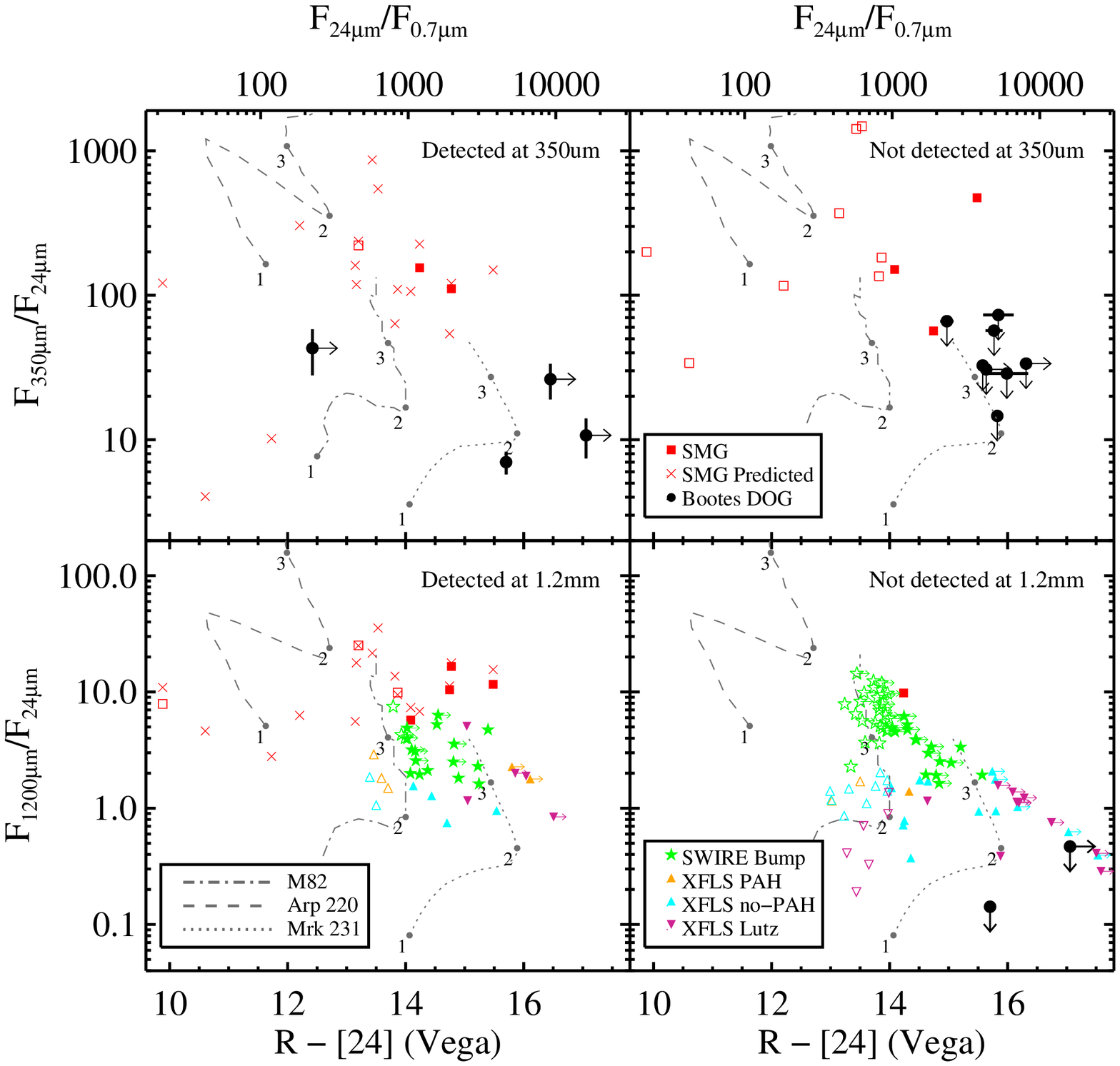}

\caption{ Top and bottom panels show 350$\mu$m/24$\mu$m and 1200$\mu$m/24$\mu$m
flux density ratios, respectively, as a function of $R$-[24] (Vega mag) color.
For clarity, sources are separated into those detected at 350$\mu$m or
1200$\mu$m on the left and those that are not detected on the right.  Objects
that qualify as a DOG ($R$ - [24] $>$ 14) are shown with a filled symbol.
Black circles indicate DOGs in Bo\"otes.  Red squares show measurements of SMGs
\citep{2008MNRAS.384.1597C}, while red crosses show predicted values based on
850$\mu$m photometry (see text for details).  Green stars show {\it
Spitzer}-selected bump sources from SWIRE \citep{2009ApJ...692..422L}.  Orange
triangles (teal triangles) show similarly identified sources from the XFLS
dominated by PAH emission (silicate absorption) features
\citep{2008ApJ...683..659S}.  Magenta inverted triangles show XFLS sources from
\citet{2008ApJ...683..659S}.  Finally, dotted, dashed, and dot-dashed lines
show the evolution of
Mrk~231, Arp~220, and M82 in this parameter space over redshift 1 to 3.  The 
DOGs studied in this paper have some of the reddest $R-[24]$ colors and lowest
far-IR/mid-IR flux density ratios of other $z \sim 2$ ULIRGs like SMGs or other
{\it Spitzer}-selected sources.  }

\label{fig:rm24trend}

\end{figure*}

An alternative way of displaying this information is shown in
Figure~\ref{fig:rm24trend}.  In each panel, the flux density ratio
far-IR:mid-IR is plotted as a function of the flux density ratio mid-IR:optical
($F_{\rm 24\mu m}/F_{\rm 0.7\mu m}$).  The top two panels show $F_{\rm 350\mu
m} / F_{\rm 24\mu m}$ on the y-axis, while the bottom two panels show $F_{\rm
1200\mu m} / F_{\rm 24\mu m}$ on the y-axis.  SMGs and various {\it
Spitzer}-selected sources are shown in the plots, divided into those that are
detected in the (sub-)mm on the left and those that are not detected on the
right.  The SMG, XFLS, and SWIRE $R$-band data come from
\citet{2008MNRAS.386.1107D}, \citet{2007ApJ...658..778Y}, and
\citet{2009ApJ...692..422L}, respectively.  For SMGs without detections at
1200$\mu$m, $F_{\rm 1200\mu m}$ is estimated using the 850$\mu$m flux density
and the dust temperature from \citet{2008MNRAS.384.1597C}, and is represented
by a red cross symbol.  Dotted, dashed, and dot-dashed lines indicate the
evolution of Mrk~231, Arp~220, and M82, respectively, on this diagram over
redshifts of $1-3$.  Compared to SMGs, DOGs in this sample have redder flux
density ratios in the mid-IR:optical but bluer far-IR:mid-IR ratios.  This
cannot be explained by an enhancement of the 24$\mu$m flux due to PAH emission,
since mid-IR spectra of these DOGs show power-law continua with silicate
absorption and weak or absent PAH emission features
\citep{2005ApJ...622L.105H}.  Instead, the most likely explanation is that
obscured AGN emission boosts the mid-IR continuum
\citep[e.g.,][]{1981ApJ...250...87R} relative to both the optical and far-IR.

\subsection{IR Luminosities}\label{sec:irlum}

In this section, we provide the best available estimates of the total IR
luminosities ($L_{\rm IR}$; 8-1000$\mu$m rest-frame) of the sample in this
paper based on the 350$\mu$m imaging.  We then compare these with estimates
based solely on the 24$\mu$m flux density, and also with estimates of the
far-IR luminosity ($L_{\rm FIR}$; 40-500$\mu$m, rest-frame) based on a modified
black-body which has been scaled to match the sub-mm photometry. 

The qualitative SED comparison from section~\ref{sec:seds} suggests that
Mrk~231 provides a reasonable fit to the far-IR photometry.  In addition,
analysis of 70$\mu$m and 160$\mu$m photometry of a sample of these types of
AGN-dominated DOGs has suggested that Mrk~231 provides a reasonable
approximation of the full SED \citep[][see table~\ref{tab:photometry} for
overlap between that study and this one]{2009ApJ...691.1846T}.  Therefore, as
the best measure of $L_{\rm IR}$, we integrate (over 8-1000$\mu$m rest-frame) a
redshifted Mrk~231 template which has been scaled to match the observed
350$\mu$m flux density (or 3-$\sigma$ limit, in the case of a non-detection).
These values are tabulated in the first column of Table~\ref{tab:luminosities}.
For sources with detections at 350$\mu$m, $L_{\rm IR}$ is in the range of
$(2.0-2.6)\times10^{13} \: L_\sun$, with a median value of $2.2\times10^{13} \:
L_\sun$.

The second column of Table~\ref{tab:luminosities} shows the rest-frame 8$\mu$m
luminosity, $\nu L_\nu (8\mu{\rm m})$, while the third column shows the total
IR luminosity (8-1000$\mu$m, rest-frame) based on the spectroscopic redshift
and an empirically determined relationship between $\nu L_\nu (8\mu{\rm m})$
and $L_{\rm IR}$: $L_{\rm IR} = 1.91L_8^{1.06}$ \citep{2007ApJ...660...97C}.
This approach was used by \citet{2008ApJ...677..943D} in determining the
contribution of DOGs to the total IR luminosity density of all $z \sim 2$
galaxies.  $L_{\rm IR}$ values range over (0.5-8.1)~$\times$~10$^{13} \:
L_\sun$, with a median value of 2.3~$\times$~10$^{13} \: L_\sun$.  The Caputi
et al. derived value for $L_{\rm IR}$ is consistent with the 350$\mu$m-based
estimate (or 3-$\sigma$ limit, in the case of non-detections) in 6/12 of the
power-law DOGs studied here.  In the remaining half of the sample, the Caputi
et al. relation overestimates $L_{\rm IR}$ in 5/6 targets.  In only one DOG
(S8) is the 350$\mu$m emission brighter than would be expected based on the
24$\mu$m flux density, redshift, and the Caputi et al. relation.  This implies
that measurements of the IR luminosity density of DOGs relying solely on the
24$\mu$m flux density will tend to overestimate their true contribution,
consistent with what has been found in a recent study of faint ($F_{\rm 24\mu
m}\sim 100-500\mu$Jy) DOGs in GOODS-N \citep{2008arXiv0808.2816P}.  Quantifying
the extent of this effect will require much larger samples of DOGs with sub-mm
measurements, the kind that will result from wide field surveys with {\it
Herschel} and SCUBA-2.

Finally, the last column of Table~\ref{tab:luminosities} shows FIR luminosities
computed from the integral over 40-500$\mu$m (rest-frame) of the best-fit
modified black-body (described in more detail in section~\ref{sec:dusttemp}).
These values are tabulated only for those sources with CARMA 1mm imaging.  We
find $L_{\rm FIR}$ values of $\approx 10^{13} \: L_\sun$, implying $L_{\rm
IR}/L_{\rm FIR} \approx 3$.  In contrast, Mrk~231 has $L_{\rm IR}/L_{\rm FIR}
\approx 2$, underscoring the fact that the IR luminosity of these DOGs is
dominated by mid-IR emission rather than FIR emission.


\begin{deluxetable*}{lllll}
\tabletypesize{\scriptsize}
\tablecolumns{5}
\tablecaption{Luminosities}
\tablehead{
\colhead{} &
\colhead{$L_{\rm IR}$\tablenotemark{a}} &
\colhead{$\nu L_\nu (8{\rm \mu m})$} &
\colhead{$L_{\rm IR}$\tablenotemark{b}} &
\colhead{$L_{\rm FIR}$\tablenotemark{d}} \\
\colhead{Source Name} &
\colhead{($10^{12} \: L_\sun$)} &
\colhead{($10^{12} \: L_\sun$)} &
\colhead{($10^{12} \: L_\sun$)} &
\colhead{($10^{12} \: L_\sun$)} 
}
\startdata
S1    &   $<32$\tablenotemark{e}     & 2.2 &  23& ---         \\
S2    &   $26\pm5$\tablenotemark{d}  & 2.4 &  25& 9.1  \\
S3    &   $22\pm7$\tablenotemark{d}  & 7.2 &  81& 10  \\
S4    &   $<44$\tablenotemark{e}     & 5.2 &  57& ---         \\
S5    &   $<40$\tablenotemark{e}     & 53  & 5.1& ---         \\
S6    &   $<64$\tablenotemark{e}     & 1.5 &  15& ---         \\
S7    &   $21\pm 7$\tablenotemark{d} & 2.1 &  22 & ---         \\
S8    &   $20\pm 5$\tablenotemark{d} & 1.1 &  11 & ---         \\
S9    &   $<75$\tablenotemark{e}     & 2.9 &  31& ---         \\
S10   &   $<28$\tablenotemark{e}     & 6.2 &  69& ---          \\
S11   &   $<33$\tablenotemark{e}     & 3.9 &  42& ---          \\
S12   &   $<16$\tablenotemark{e}     & 2.0 &  21& ---          \\
\tablenotetext{a}{Integral over 8-1000$\mu$m of redshifted Mrk~231 template normalized at 350$\mu$m.}
\tablenotetext{b}{Estimated from $\nu L_\nu$~(8$\mu$m) - $L_{\rm IR}$ relation
from \citet{2007ApJ...660...97C}.}
\tablenotetext{c}{Integral over 40-500$\mu$m of best-fit modified black-body
(only sources with CARMA 1mm data).}
\tablenotetext{d}{Uncertainties shown reflect 350$\mu$m photometric
uncertainties.  Addtional systematic uncertainties associated with the adoption
of a Mrk~231 template are not included.}
\tablenotetext{e}{3$\sigma$ upper limits.}
\enddata
\label{tab:luminosities}
\end{deluxetable*}

\subsection{Constraints on Dust Properties}\label{sec:dusttemp}

Additional constraints can be placed on the nature of the cold dust emission
from the two sources (S2 and S3) with CARMA.  The 1mm non-detections imply warm
dust temperatures.  If we compute the predicted flux density at 1mm based on
the observed 24$\mu$m flux density and assuming the three local galaxy SED
templates (M82, Arp~220, and Mrk~231) used in the previous sections, we find
values of 4.4, 740, 1.8~mJy, and 5.8, 120, 2.7~mJy for the two sources
respectively.  For S2, this implies that our 3-$\sigma$ limit ($<1.5 \:$mJy) is
close to the level we would expect if Mrk~231 is an appropriate SED, while the
other two SEDs are clearly inconsistent with the data.  For S3, the limit
($<1.8 \:$mJy) is inconsistent with each of the SEDs, implying a warmer dust
temperature than even Mrk~231.

A more quantitative approach is to minimize the residuals between the
160$\mu$m, 350$\mu$m, and 1mm data and that expected from a modified black
body.  Doing this, we can constrain the dust temperature, $T_{\rm dust}$, and
the dust emissivity index, $\beta$, of each target.  The best-fit quantities
and their uncertainties are estimated using a bootstrap technique that mimics
the procedure used by \citet{2000MNRAS.315..115D}.  Briefly, for each flux
density measurement, a set of 100 artificial flux densities are generated using
a Gaussian random number generator.  The mean value and dispersion of the
distribution of artificial flux densities are set by the measurement value and
its 1$\sigma$ uncertainty, respectively (for non-detections, we assume a mean
value of 0 and force artifical flux densities to be positive).  Each artificial
SED is used to construct a distribution of best-fit scaling factors and
associated $\chi^2_\nu$ values for a modified black-body with a given
combination of $\beta$ and $T_{\rm dust}$.

\begin{figure}[!tbp]
\epsscale{1.0}
\plotone{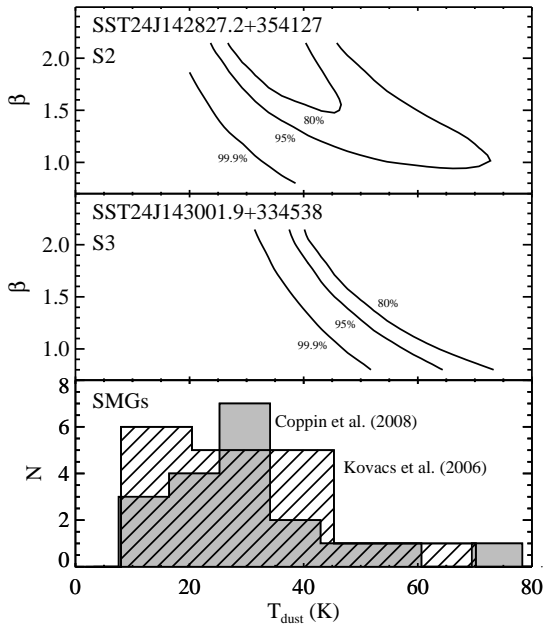}

\caption{$\chi^2_\nu$ contours based on modified black-body fits to sources
with CARMA data.  Lines indicate 80\%, 95\%, and 99.9\% confidence intervals.
Also shown are $T_{\rm dust}$ distributions for SMGs with 350$\mu$m data
\citep{2006ApJ...650..592K,2008MNRAS.384.1597C}.  The 95\% confidence levels
for S2 and S3 suggest dust temperature limits that would place them in the
warmest 50\% and 15\%, respectively, of SMGs. }

\label{fig:tdbeta}

\end{figure}

Figure~\ref{fig:tdbeta} shows the median $\chi^2_\nu$ contours for the grid of
$T_{\rm dust}$ and $\beta$ values we have sampled in the model fitting for each
source observed by CARMA.  The contours show the degeneracy between $\beta$ and
$T_{\rm dust}$, and illustrate why a perfect fit is not possible in spite of
the fact that a model with three parameters is being fit to three data points
(i.e., the input parameters are not fully independent).  The distribution of
$T_{\rm dust}$ found for SMGs based on 350$\mu$m imaging from
\citet{2008MNRAS.384.1597C} and \citet{2006ApJ...650..592K} is displayed in the
lower panel.  Two-sided Kolmogorov-Smirnov tests suggest that the distribution
of $T_{\rm dust}$ (assuming $\beta=1.5$) for each CARMA target is highly
unlikely to be drawn from the same parent distribution as the combined sample
of SMGs (4\% and 0.008\%, for the two CARMA targets respectively).  In both of
the CARMA targets, warmer dust temperatures are required due to the
non-detection at 1mm.  The data cannot rule out models with higher values of
$\beta$, but models of dust grains as well as Galactic and extra-galactic
observations consistently suggest $\beta = 1 - 2$
\citep{1983QJRAS..24..267H,2001MNRAS.327..697D}.  For $\beta=1.5$, we reject
$T_{\rm dust}\leq 33$K and 45K at the 95\% confidence level, in S2 and S3
respectively. For $\beta=2$, we can reject $T_{\rm dust}\leq 25$~K and 37~K.

\subsection{Dust Masses}\label{sec:masses}

Assuming optically thin sub-mm emission, cold dust masses can be estimated from
the 350$\mu$m photometry \citep{1997MNRAS.289..766H}:

\begin{equation}
M_{\rm dust} = \frac{1}{1+z} \frac{S_{\rm obs} D_L^2}{\kappa_d^{\rm rest}
B(\nu^{\rm rest}, \: T_{\rm dust})}
\label{eq:mdust}
\end{equation}

\noindent where $S_{\rm obs}$ is the observed 350$\mu$m flux density, and
$\kappa_d^{\rm rest}$ and $B(\nu^{\rm rest}, T_{\rm dust})$ are, respectively,
the values of the mass absorption coefficient and black-body function at the
rest frequency $\nu^{\rm rest}$ and dust temperature $T_{\rm dust}$.  The
appropriate value for $\kappa_d$ is uncertain to at least a factor of two
\citep{2003MNRAS.341..589D}; we use a $\kappa_d^{\rm rest}$ value interpolated
from \citet{2003ARA&A..41..241D} ($<$$\kappa_d^{\rm rest}$$> \approx 20
\:$cm$^2\:$g$^{-1}$).  

The results from section~\ref{sec:dusttemp} suggest that $T_{\rm dust} >
35-60$~K for two of the sources.  Adopting the average of these limits ($T_{\rm
dust}=45~$K) for all of the sources in this sample, then dust mass limits are
in the range ($4.1-7.9$)~$\times$~10$^{8} \: M_\sun$ (median value of
5.1$\times$10$^{8} \: M_\sun$) for the five objects with detections at
350$\mu$m.  The 3$\sigma$ upper limits on the dust masses of the remaining
sample range from ($3-15$)~$\times$~10$^8\:M_\sun$.  Warmer values of $T_{\rm
dust}$ would lead to smaller inferred dust masses (e.g., increasing the dust
temperature by 10~K implies $\approx$50\% lower dust masses).  The dust masses
are presented in Table~\ref{tab:masses}.  

These values agree with those of \citet{2009ApJ...693..750B}, where dust masses
of a sample of 31 power-law dominated DOGs were estimated using predicted
850$\mu$m flux densities based on Mrk~231 templates and the measured 24$\mu$m
flux density.  In previous work, we assumed a $T_{\rm dust}$ of 75~K and found
a median dust mass of 1.6$\times$10$^8 \: M_\sun$.  This is consistent with the
notion that Mrk~231 accurately characterizes the far-IR SED of power-law DOGs,
as described in section~\ref{sec:seds} and in \citet{2009ApJ...691.1846T}.

Finally, assuming a gas mass to dust mass ratio of 120 \citep[as was found in a
study of the nuclear regions of nearby LIRGs; see][]{2008arXiv0806.3002W}, then
the gas masses can be estimated as well.  Using the assumed ratio, we find gas
masses of ($5-10$)~$\times$~10$^{10} \: M_\sun$ (median value of
6$\times$10$^{10} \: M_\sun$) for the detected objects and gas mass 3$\sigma$
limits of ($4-18$)~$\times$~10$^{10}\:M_\sun$ in the remaining sample.  We
caution that this is very uncertain; \citet{2006ApJ...650..592K} report a gas
mass to dust mass ratio of $\approx 60$ for SMGs, assuming $\kappa_d^{\rm rest}
= 15\:$cm$^2\:$g$^{-1}$.  If this gas to dust mass ratio is appropriate for our
sample, then the implied gas masses will be a factor of two lower ($2-5 \times
10^{10} \: M_\sun$)

\subsection{Stellar Masses}\label{sec:mstar}

In this section, we describe the methodology and present estimates for the
stellar masses of the DOGs in this sample.

\subsubsection{Methodology}

To estimate stellar masses, we rely on Simple Stellar Population (SSP) template
SEDs from the \citet{2003MNRAS.344.1000B} population synthesis library.  All
models used here have ages spaced logarithmically from 10~Myr up to 1~Gyr,
solar metallicity, a Chabrier initial mass function (IMF) over the mass range
$0.1-100 \: M_\sun$ \citep{2003PASP..115..763C}, and use the Padova 1994
evolutionary tracks \citep{1996A&AS..117..113G}.  The reddening law from
\citet{2000ApJ...533..682C} is used between $0.12 - 2.2 \: \mu$m and that of
\citet{2003ARA&A..41..241D} for longer wavelengths.  This method is similar to
that used in \citet{2009ApJ...693..750B}.   

For sources at $z \sim 2$ whose mid-IR luminosity is dominated by stellar
light, IRAC photometry samples the SED over the wavelength range where emission
from asymptotic and red giant branch stars as well as low-mass main-sequence
stars produces an emission peak at rest-frame 1.6$\mu$m.  In such cases, for
given assumptions regarding the star-formation history, metallicity, and IMF,
stellar mass estimates can be obtained via stellar population synthesis
modeling.  One goal of this work is to estimate stellar masses using
self-consistent modeling of photometry measured at similar wavelengths for a
variety of $z \sim 2$ dusty galaxies.  Therefore, we apply this method to
determine stellar masses in SMGs as well as XFLS and SWIRE sources.  The IRAC
data for each of these galaxy populations comes from, respectively,
\citet{2008MNRAS.386.1107D}, \citet{2005ApJS..161...41L}, and
\citet{2009ApJ...692..422L}.

The DOGs studied in this paper have mid-IR SEDs that are dominated by a
power-law component, suggesting that obscured AGN emission is overwhelming the
stellar flux at these wavelengths.  The shape of the mid-IR SED therefore
provides limited constraints on the stellar population and additional
information is needed to estimate the stellar mass of these sources.  To
overcome this challenge, SSP models were fit to high spatial-resolution {\it
HST} photometry in the rest-frame UV (WFPC2/F606W or ACS/F814W) and rest-frame
optical (NIC2/F160W) from \citet{2009ApJ...693..750B}.  Two sources currently
lack {\it HST} data (SST24~J142827.2+354127 and
SST24~J143411.0+331733\footnote{ {\it HST} data exist for this source but will
be presented in a separate paper (Prescott et al., in prep.)}) and so are
excluded from this analysis.  

In principle, rest-frame near-IR data offer a means to estimate the SSP age and
$A_V$ independently, since rest-frame optical and near-IR photometry sample the
SED above the 4000~\AA\ break while the rest-frame UV photometry samples galaxy
light below the 4000~\AA\ break.  However, these data come from the IRAC
3.6$\mu$m images of the Bo\"otes field ({\it Spitzer} Deep Wide-Field Survey;
Ashby et al., in prep.), where the spatial resolution is insufficient to
resolve the nuclear source from the extended galaxy component.  In this case,
there are only limited constraints on the amount of non-stellar (i.e., obscured
AGN) emission at 3.6$\mu$m.  We have explored the effect of this uncertainty on
the fitting process by artificially reducing the 3.6$\mu$m flux by 50\%
(corresponding to the situation where the 3.6$\mu$m emission is equal parts
starlight and AGN) and re-analyzing the data.  Comparing results, we find that
higher AGN fractions imply younger inferred SSP ages and higher $A_V$ values.
Because the AGN fraction in these sources is currently unknown at 3.6$\mu$m, we
are unable to constrain both the age and $A_V$ independently.


Although the usage of photometry at different wavelengths is not ideal for the
purposes of comparing stellar masses between galaxy populations, this method
remains valuable because the models being fit to the data are the same for each
galaxy population.  Indeed, recent work has suggested that the dominant source
of systematic uncertainty in stellar mass estimates of $K$-selected galaxies at
$z \sim 2.3$ is the use of different stellar population synthesis codes
\citep{2009arXiv0906.2012M}, and that these systematics often dominate the
formal random uncertainty.  As long as the parameters of the model used here
(such as the IMF, star-formation history, metallicity, etc.) do not vary from
population to population, then the comparison presented here should be valid in
a global sense.

\subsubsection{Stellar Mass Estimates}

\begin{figure*}[!tbp]
\epsscale{1.0}
\plotone{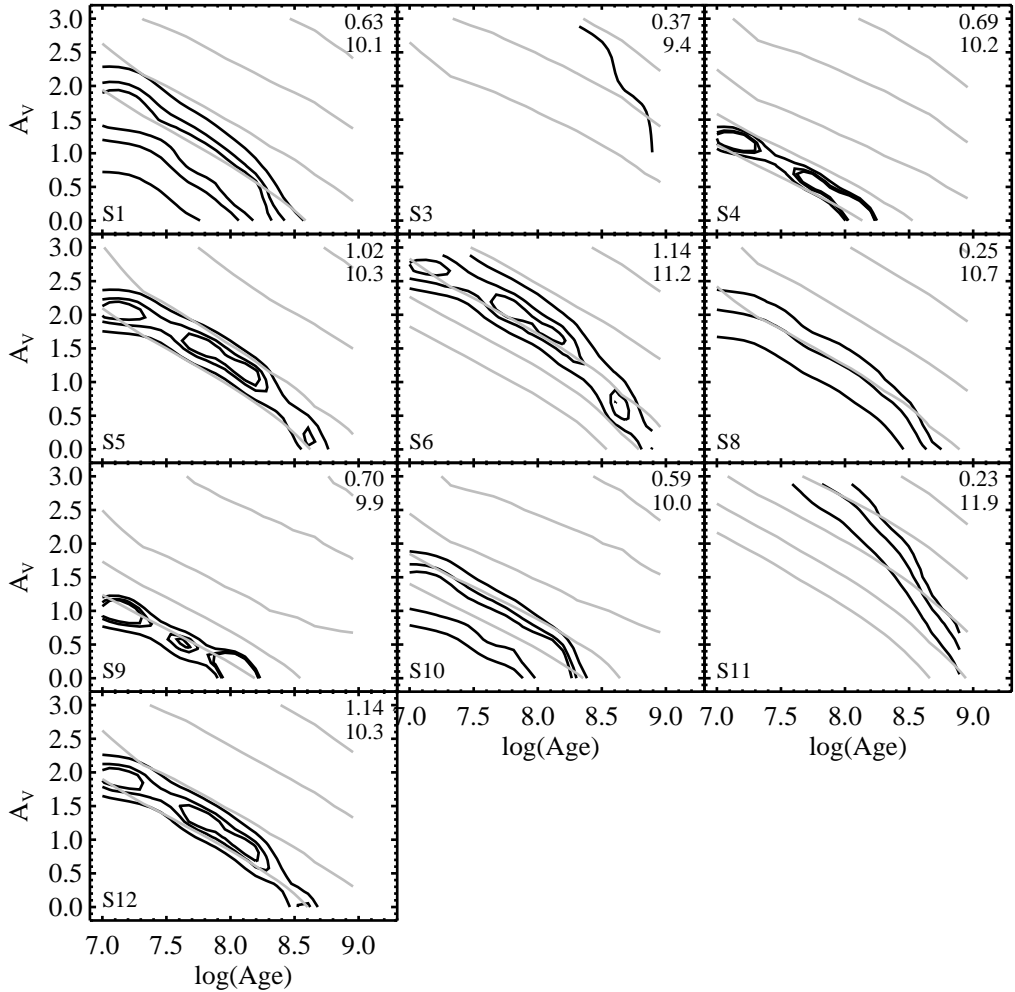}

\caption{$\chi^2_\nu$ contours based on SSP fits to {\it HST} imaging in
$H$-band and $V$- or $I$-band.  Black lines indicate 80\%, 95\%, and 99.9\%
confidence intervals.  Grey contours trace lines of constant stellar mass for
the given {\it HST} photometry, starting with $M_{\rm star} = 10^{10} \:
M_\sun$ in the bottom left and increasing by 0.5 dex towards the upper right.
In the top right corner of each panel is the minimum $\chi^2_\nu$ value and the
associated stellar mass, in units of log($M_{\rm star}/M_\sun$).  The bottom
left corner contains the source identifier.  Not shown are sources S2 and S7,
since these targets have no {\it HST} imaging available.  The best-fit stellar
masses range from $10^{10} - 10^{12} \: M_\sun$.}

\label{fig:mstar_SED}

\end{figure*}

Figure~\ref{fig:mstar_SED} shows $\chi^2_\nu$ contours for a grid of SSP ages
and $A_V$ values.  The contours trace lines of 80\%, 95\%, and 99.9\%
confidence intervals allowed by the photometric uncertainties (estimated using
a bootstrap method similar to that outlined in section~\ref{sec:dusttemp}).
The solid grey lines trace iso-mass contours and show the range in stellar
masses allowed by the photometric uncertainties.  The best-fit SSP model
parameters ($M_{\rm star}$ and $\chi^2_\nu$) are printed in each panel and
shown in Table~\ref{tab:masses}.  

The stellar masses in the sample range from $(1-20) \times 10^{10} \: M_\sun$,
with a median value of $2 \times 10^{10} \: M_\sun$.  
The $\chi^2_\nu$ values range from $0.37 - 1.16$, with a median value of 0.69.  
In one case (SST24~J143001.9+334538, or S3), the photometric uncertainty is so
large that $\chi^2_\nu < 1$ over the full range of $A_V$ and SSP age that we
have sampled and so the range of acceptable fits is very large.  For this
source, we quote the 3$\sigma$ upper limit on the stellar mass based on the
photometric uncertainty.

The ratio of the stellar to gas mass, $\zeta\equiv M_{\rm star}/M_{\rm gas}$,
is a measure of the evolutionary state of the galaxy, with larger values
indicating more processing of gas to stars. Our estimates of $\zeta$, computed
assuming $M_{\rm gas}/M_{\rm dust}=120$, are presented in Table 5.  We caution
that the gas mass to dust mass ratio is highly uncertain.  In SMGs, there is
evidence suggesting it is $\approx 60$ \citep{2006ApJ...650..592K}.  Adopting
this lower value would imply lower gas masses by a factor of two and hence
double our $M_{\rm star}/M_{\rm gas}$ estimates.


Clustering studies suggest that the most luminous DOGs reside in very massive
halos \citep[$M_{\rm DM} \sim 10^{13} \: M_\sun$][]{2008ApJ...687L..65B}. It is
tempting to attribute the low stellar masses we estimate for DOGs to youth.
However, the absolute stellar masses we compute are extremely uncertain.  For
example, the use of a Salpeter IMF rather than a Chabrier IMF would
approximately double our stellar mass estimates \citep{2003MNRAS.344.1000B}.
Beyond the choice of what IMF slope to use, the mass-to-light ratio of a model
galaxy (for a given rest-frame near-UV - $R$ color) can vary significantly
depending on the details of its star-formation history, the clumpiness of its
interstellar medium and the associated dust attenuation law, as well as how
advanced stages of stellar evolution are treated, such as blue stragglers,
thermally-pulsating asymptotic giant branch stars, etc.
\citep{2009arXiv0904.0002C}.  In light of these uncertainties, the fact that
our stellar mass estimates are low ($M_{\rm star} \sim 10^{10} \: M_\sun$)
compared to the dark matter haloes in which we believe they reside is not yet a
cause for concern -- a quantitative study of the maximum possible stellar mass
allowed by the photometry (by examining results from different stellar
population synthesis codes, star-formation histories, metallicities, etc.)
would be the best way to approach this issue in the near-term, but is beyond
the scope of the current work.

\begin{deluxetable}{lrrr}
    \tabletypesize{\small}
    \tablecolumns{4}
    \tablecaption{Dust Masses and Stellar Properties}
    \tablehead{
    \colhead{} &
    \colhead{$M_{\rm dust}$\tablenotemark{a}} &
    \colhead{$M_{\rm star}$\tablenotemark{b}} &
    \colhead{} \\
    \colhead{Source Name} &
    \colhead{($10^8 \: M_\sun$)} &
    \colhead{($10^{10} \: M_\sun$)} &
    \colhead{$M_{\rm star}/M_{\rm gas}$}
    }
    \startdata
S1   &    $<6.3$ &    $(0.3-1.5) $ & $0.05 - 0.24$ \\
S2   &     $7.9\pm1.4$ &  --- & --- \\     
S3   &     $4.9\pm1.5$ &  $  <10  $ &  $<1.7$ \\
S4   &    $<5.5$ &  $(1-3       )$ & $>0.15 - 0.45$ \\
S5   &    $<6.2$ &  $(1.5-3)     $ & $>0.20 - 0.40$ \\
S6   &    $<7.5$ &  $(10-20)       $ & $>1.1 - 2.2$ \\
S7   &     $2.7\pm0.7$ &  --- & --- \\     
S8   &     $2.5\pm0.9$ &  $ > 1        $ & $>0.33$ \\
S9   &    $<8.5$ &  $(0.6 - 1.1) $ & $>0.06 - 0.11$ \\
S10  &    $<6.5$ &  $(0.5 - 3)   $ & $>0.06 - 0.38$ \\
S11  &    $<4.0$ &  $(  > 10  )   $ & $>2.1$ \\
S12  &    $<1.9$ &  $( 1 - 2 )   $ & $>0.44 - 0.88$ \\
    \tablenotetext{a}{Dust mass assuming $T_{\rm dust} = 45$~K.}
    \tablenotetext{b}{Stellar mass estimated from fitting photometry in the
    rest-frame UV and optical ($V/I$ and $H$ respectively).  Range given
    reflects 95\% confidence intervals based only on photometric uncertainty.}
    \enddata
    \label{tab:masses}
\end{deluxetable}

\section{Discussion}\label{sec:disc}

In this section, we seek to understand the role of DOGs in galaxy evolution and
their relation to other high-$z$ galaxy populations.  We begin by motivating a
comparison sample of such objects, including SMGs and {\it Spitzer}-selected
ULIRGs from the XFLS and SWIRE survey.  We then examine how the measured
properties differ from population to population.  We end with the implications
of these comparisons for models of galaxy evolution. 

\subsection{Related $z \approx 2$ Galaxy Populations}\label{sec:populations}

\subsubsection{SMGs}\label{sec:smgs}

SMGs represent an interesting population of galaxies for comparison with DOGs
because they are selected at sub-mm wavelengths where the dominant emission
component is cold dust ($T_{\rm dust} \sim 30~$K).  In contrast, DOGs are
selected predominantly by their brightness at 24$\mu$m and therefore should be
dominated by hot dust.  Despite this fundamental distinction, these two galaxy
populations have similar number densities and redshift distributions
\citep{2005ApJ...622..772C,2008ApJ...677..943D,2004ApJ...611..725B}.  Recent
evidence suggests that 24$\mu$m-faint ($F_{\rm24\mu m}\sim 0.1-0.5$~mJy) DOGs
have a composite SED whose shape in the far-IR closely mimics that of the
average bright ($F_{850} > 5~$mJy) SMG \citep{2008arXiv0808.2816P}.
Futhermore, 24$\mu$m-faint DOGs and SMGs have similar real space correlation
lengths ($r_0 \approx 6 \pm 2 ~ h^{-1} \:$Mpc), yet there is tentative evidence
that DOG clustering strength increases with 24$\mu$m flux density \citep[$r_0
\approx 13 \pm 3 ~ h^{-1} \:$Mpc for DOGs with $F_{\rm24\mu m} >
0.6$~mJy;][]{2008ApJ...687L..65B}.  While these results are suggestive of an
association between the two populations, the details of such a connection are
not yet clear.  In an effort to study this connection via their far-IR
properties, we will compare the data presented in this paper with SHARC-II
350$\mu$m and MAMBO 1.2mm imaging of 25 SMGs from the Submillimetre Common User
Bolometer Array (SCUBA) HAlf Degree Extragalactic Survey
\citep{2006ApJ...643...38L,2006ApJ...650..592K,2008MNRAS.384.1597C,2004MNRAS.354..779G}.

\subsubsection{XFLS Sources}\label{sec:xfls}

A set of {\it Spitzer}-selected galaxies from the 4 deg$^2$ XFLS share many
properties with the 24$\mu$m-bright DOGs \citep{2007ApJ...658..778Y}.  The
specific selection criteria are similar, although not necessarily as extreme in
their IR-optical flux density ratios: $F_{24\mu{\rm m}} \geq 0.9$~mJy, $\nu
F_\nu(24\mu{\rm m})/\nu F_\nu(8\mu{\rm m}) \geq 3.16$, and $\nu F_\nu(24\mu{\rm
m})/\nu F_\nu(0.7\mu{\rm m}) \geq 10$ (in comparison, DOGs have  $\nu
F_\nu(24\mu{\rm m})/\nu F_\nu(0.7\mu{\rm m}) \geq 30$).  {\it Spitzer}/IRS
spectroscopy of these objects has revealed strong silicate absorption and in
some cases PAH emission features on par with those of SMGs
\citep{2007ApJ...667L..17S}.  This suggests that the XFLS sources are composite
AGN/starburst systems and may represent a transition phase between (un)obscured
quasars and SMGs \citep{2008ApJ...683..659S}.  MAMBO 1.2mm observations of 44
XFLS sources have allowed a detailed study of their far-IR properties and have
suggested $<$$L_{\rm IR}$$>$ $\sim$ 7$\times$10$^{12} \: L_\sun$
\citep{2008ApJ...683..659S}.

\subsubsection{SWIRE Sources}\label{sec:swire}

The last set of comparison galaxies we consider are {\it Spitzer}-selected
sources from the SWIRE survey \citep{2009ApJ...692..422L}.  Like DOGs in
Bo\"otes and the XFLS sources, they have large IR-optical flux density ratios.
However, an additional criterion has been applied to identify sources with
significant emission at rest-frame 1.6$\mu$m due to evolved stellar
populations.  For sources at $z=1.5-3$, this means selecting objects whose
mid-IR spectrum peaks at 5.8$\mu$m.  Although spectroscopic redshifts are not
available for most of this sample, SED fitting has suggested photometric
redshifts consistent with $z \sim 2$ and stellar masses of (0.2 -
6)$\times$10$^{11} \: M_\sun$ \citep{2009ApJ...692..422L}.  MAMBO 1.2mm
photometry for 61 of these SWIRE sources has indicated far-IR luminosities of
10$^{12}$-10$^{13.3} \: L_\sun$ \citep{2009ApJ...692..422L}.

\subsection{Comparison of Measured Properties}\label{sec:properties}

Our results from section~\ref{sec:results} represent our best estimates of
$L_{\rm IR}$, $L_{\rm FIR}$, $T_{\rm dust}$, $M_{\rm dust}$, and $M_{\rm star}$
for the DOGs in the sample.  In Table~\ref{tab:comparison}, we give the median
value of these quantities for DOGs in Bo\"otes (from this paper), SMGs, and
XFLS and SWIRE sources.  In computing these median values, we do not consider
sources at $z < 1$; nor do we consider sources without detections at (sub-)mm
wavelengths (see discussion on caveats to the analysis at the end of this
section).  Table~\ref{tab:comparison} also makes a distinction between XFLS
sources whose mid-IR spectra are dominated by strong PAH features (XFLS PAH)
and those that show weak or absent PAH features (XFLS weak-PAH).  Each of these
galaxy populations is further subdivided into those that qualify as DOGs ($R -
[24] > 14$) and those that do not. 

\begin{deluxetable*}{lllllllll}
\tabletypesize{\small}
\tablecolumns{9}
\tablecaption{Average High-$z$ Galaxy Properties}
\tablehead{
\colhead{} &
\colhead{} &
\colhead{} &
\colhead{$L_{\rm IR}$} &
\colhead{$L_{\rm FIR}$} &
\colhead{$T_{\rm dust}$} &
\colhead{$M_d$} &
\colhead{$M_{\rm star}$} &
\colhead{} \\
\colhead{Source} &
\colhead{$R$-[24]} &
\colhead{$N$} &
\colhead{($10^{12} \: L_\sun$)} &
\colhead{($10^{12} \: L_\sun$)} &
\colhead{(K)} &
\colhead{($10^9 \: M_\sun$)} &
\colhead{($10^{10} \: M_\sun$)} &
\colhead{$M_{\rm star}/M_{\rm gas}$\tablenotemark{a}}
}
\startdata
Bo\"otes\tablenotemark{b}      & $>$ 14 &   5 & 23  & 10 & 45 & 0.5 & $>2$& $>0.3$ \\
\\                                                                                
    XFLS\tablenotemark{c}      &    ALL &  11 & 7.7 & 2.7 & 32 & 5.0 & 13 & 0.48  \\
                               & $>$ 14 &   6 & 8.7 & 1.1 & 27 & 7.3 & 10 & 0.21  \\
                               & $<$ 14 &   5 & 6.5 & 4.6 & 37 & 2.3 & 16 & 0.58  \\
\\                                                                                
XFLS PAH\tablenotemark{c}      &    ALL &   5 & 5.7 & 1.8 & 31 & 4.6 & 22 & 0.54  \\
                               & $>$ 14 &   2 & 6.3 & 1.6 & 29 & 7.5 & 23 & 0.30  \\
                               & $<$ 14 &   3 & 5.4 & 2.0 & 32 & 2.8 & 22 & 0.69  \\
\\                                                                                
XFLS weak-PAH\tablenotemark{c} &    ALL &   6 & 9.4 & 3.4 & 32 & 5.3 & 5.0 & 0.25 \\
                               & $>$ 14 &   4 & 3.5 & 0.8 & 26 & 7.2 & 3.8 & 0.17 \\
                               & $<$ 14 &   2 & 14 & 8.0 & 45 & 1.6 & 7.3 & 0.40 \\
\\                                                                                
   SWIRE\tablenotemark{d}      &    ALL &  19 & 6.5 & 3.1 & 32 & 6.7 & 28 & 0.53  \\
                               & $>$ 14 &  16 & 6.7 & 3.1 & 32 & 6.1 & 28 & 0.56  \\
                               & $<$ 14 &   3 & 5.5 & 2.6 & 30 & 9.9 & 24 & 0.35  \\
\\                                                                                
     SMG\tablenotemark{e}      &    ALL &  18 & 6.9 & 3.2 & 35 & 1.5 & 10 & 0.60  \\
                               & $>$ 14 &   4 & 7.8 & 3.6 & 28 & 2.9 & 6.9 & 0.28 \\
                               & $<$ 14 &  14 & 6.1 & 2.9 & 37 & 1.1 & 12 & 0.71  \\
\tablenotetext{a}{Computed using $M_{\rm gas}/M_{\rm dust} = 120$.}
\tablenotetext{b}{Includes only DOGs detected at 350$\mu$m.}
\tablenotetext{c}{XFLS sources with MAMBO 1.2mm detections \citep{2008ApJ...683..659S}.}
\tablenotetext{d}{SWIRE sources with MAMBO 1.2mm detections
\citep{2009ApJ...692..422L}.}
\tablenotetext{e}{From compilation of \citet{2008MNRAS.384.1597C}.}
\enddata
\label{tab:comparison}
\end{deluxetable*}

The primary feature of this comparison is that the relative uncertainty in the
estimated parameters between galaxy populations has been reduced by computing
the respective values self-consistently with the methods outlined in
section~\ref{sec:results}.  The exceptions to this rule are $L_{\rm IR}$ and
$T_{\rm dust}$ (note that while the photometry used to determine $M_{\rm star}$
for Bo\"otes DOGs is different than for the other galaxy populations, the
methodology used is the same, including the use of the same set of model SSP
templates).  Our method of computing $L_{\rm IR}$ relies on the assumption that
Mrk~231 represents a reasonable approximation of the source SED.  For many SMGs
as well as XFLS and SWIRE sources, this is an unrealistic assumption.  Instead,
we estimate $L_{\rm IR}$ from $L_{\rm FIR}$, assuming (1) $L_{\rm IR} = L_{\rm
IR, \: SB} + L_{\rm IR, \: AGN}$; (2) $L_{\rm IR, \: SB} = \alpha L_{\rm FIR}$;
(3) $L_{\rm IR, \: AGN} = \epsilon L_{\rm IR}$, where $\alpha$ is a factor
$\approx 1.3$, depending on $T_{\rm d}$ and $\beta$ \citep{1988ApJS...68..151H}
and $\epsilon$ is the typical AGN fraction of the galaxy population.  For SMGs,
we adopt the conservative upper limit from \citet{2008ApJ...675.1171P} of
$\epsilon = 0.3$.  For SWIRE sources and XFLS PAH sources, this fraction is
$\approx0.3$
\citep{2008ApJ...675.1171P,2009ApJ...692..422L,2008ApJ...683..659S}, while for
XFLS weak-PAH sources we use 0.7 \citep{2008ApJ...683..659S}.  

The $T_{\rm dust}$ values given in the literature are adopted for each source.
It should be noted that $T_{\rm dust}$ for SWIRE sources are uncertain due to
the lack of data near the far-IR peak (i.e., observed-frame 160 or 350 $\mu$m).
\citet{2009ApJ...692..422L} analyze the stacked signal at 160$\mu$m from these
sources and find that the $T_{\rm dust}$ is higher by as much as 10~K than what
is assumed in their Table~\ref{tab:comparison}.  For a given 1.2mm flux,
increasing $T_{\rm d}$ by 10~K will increase $L_{\rm FIR}$ by a factor of
$\approx$3 and decrease $M_{\rm dust}$ by $\approx$50\%.  

The key result from Table~\ref{tab:comparison} is that while the DOGs in our
sample have lower dust masses than the other galaxy populations by a factor of
$\sim3-20$, they have higher total IR and far-IR luminosities by factors of
$\sim$2.  This distinction is driven by the difference in $T_{\rm dust}$, as
DOGs in the sample have higher values by $\approx$10-20~K compared to the other
galaxy populations.  

In terms of the stellar and gas mass estimates, the relationship between DOGs
in Bo\"otes (i.e., the sample studied in this paper) and the remaining galaxy
populations is unclear.  Even if a single dust temperature and a single dust to
gas mass ratio for each of the sources studied in this paper is adopted, the
uncertainties on the stellar mass estimates are large enough to allow greatly
varying stellar mass to gas mass ratios.  Sources satisfying $R-[24]>14$ (i.e.,
DOGs) tend to have higher gas masses compared to $R-[24]<14$ sources (under the
assumption of a constant dust-to-gas mass ratio).  This difference is at least
in part due to a difference in dust temperatures; within this sample, DOGs have
lower dust temperatures than non-DOGs.  This is in contrast with the evidence
for high dust temperatures in the Bo\"otes DOGs studied in this paper and may
be an indication that mm-detected DOGs represent a special subset of DOGs that
is more representative of the mm-selected galaxy population than the DOG
population.  

An important caveat to this comparison is that we are dealing with small sample
sizes due to incomplete coverage at one or more bands from the mid- to the
far-IR.  For instance, while every DOG has a measured 24$\mu$m flux density,
very few have been observed at 350$\mu$m, and only two have been observed at
1mm.  Similarly, few XFLS and SWIRE sources have been detected at 1mm and even
fewer have been observed at 350$\mu$m.  Although SMGs are the best-studied class
of objects within this set of populations, they too suffer from low-number
statistics.  Larger sample sizes in the critical 200-500$\mu$m regime will
arrive following the analysis of wide-field survey data from the Balloon-borne
Large Aperture Submillimeter Telescope \citep[e.g.,][]{2009arXiv0904.1206P} and
the {\it Herschel} Space Observatory.

\subsection{Implications for Models of Galaxy Evolution}\label{sec:implications}

One of the major open questions in galaxy evolution is the effect that AGN have
on their host galaxies.  In the local universe, there is observational evidence
that ULIRGs dominated by warm dust serve as a transition phase between cold
dust ULIRGs and optically luminous quasars and that this transition may be
driven by a major merger \citep{1988ApJ...325...74S,1988ApJ...328L..35S}.
Recent theoretical models of quasar evolution based on numerical simulations of
major mergers between gas-rich spirals have suggested that the subsequent
growth of a super-massive black hole can regulate star-formation via a feedback
effect which re-injects energy into the interstellar medium (ISM) and expels
the remaining gas that would otherwise form stars \citep{2006ApJS..163....1H}.  

Although the notion that local ULIRGs are associated with mergers is well
accepted \citep[e.g.,][]{1996ARA&A..34..749S}, the picture is less clear at
high redshift.  Morphological studies of high-$z$ galaxies suffer from surface
brightness dimming, making the detection of faint merger remnant signatures
difficult
\citep[e.g.,][]{2008ApJ...680..232D,2008AJ....136.1110M,2009ApJ...693..750B,2009AJ....137.4854M}.
However recent theoretical work on the cosmological role of mergers in the
formation of quasars and spheroid galaxies suggests that they dominate the $z
\gtrsim 1$ quasar luminosity density compared to secular processes such as bars
or disk instabilities \citep{2008ApJS..175..356H}.  

If major mergers drive the formation of massive galaxies at high redshift, then
one possible interpretation of our results involves an evolutionary scenario in
which these sources represent a very brief but luminous episode of extreme AGN
growth just prior to the quenching of star formation.  In such a scenario, SMGs
and the brightest 24$\mu$m-selected sources represent the beginning and end
stages, respectively, of the high star-formation rate, high IR luminosity phase
in massive galaxy evolution.  Consistent with this scenario is that we find
24$\mu$m-bright DOGs in Bo\"otes to have higher dust temperatures (possibly
from AGN heating of the dust) than SMGs and less extreme {\it Spitzer}-selected
sources.  
We caution, however, that these results are consistent with any evolutionary
model in which the 24$\mu$m-bright phase follows the sub-mm bright phase, be it
driven by major mergers, minor mergers, or some secular process.  

Finally, we stress that larger samples of mm and sub-mm imaging of {\it
Spitzer}-selected galaxies are needed in order to understand their role in
galaxy evolution fully by comparing samples of similar number density,
clustering properties, etc.  Much of this will be provided by upcoming {\it
Herschel} and SCUBA-2 surveys.  In the more immediate future, 1mm imaging with
currently available instruments such as the Astronomical Thermal Emission
Camera (AzTEC) and the MAx-Planck Millimeter BOlometer Array (MAMBO) will be
critical to constraining the cold dust properties of {\it Spitzer}-selected
galaxies.  Only when these surveys have obtained statistically significant
numbers of detections or stringent upper limits will we be able to make
definitive conclusions regarding the nature of the link between AGN and
starbursts in the formation of the most massive galaxies.

\section{Conclusions}\label{sec:conc}

We present CSO/SHARC-II 350$\mu$m and CARMA 1mm photometry of DOGs in the
Bo\"otes Field.  The major results and conclusions from this study are the following:

\begin{enumerate}

\item At 350$\mu$m, 4/5 DOGs are detected in data with low rms levels ($ \leq
    15 \: $mJy) and 0/8 DOGs are detected in data with medium to high rms
    levels ($20-50\:$mJy).  At 1mm, a subset of two DOGs were observed but not
    detected.

\item Mrk~231 is confirmed as a valid template for the SEDs of the DOGs in this
sample.  This suggests the 24$\mu$m bright ($F_{\rm 24\mu m} \gtrsim 1$~mJy)
population of DOGs is dominated by warm dust, possibly heated by an AGN.  Cold
dust templates such as Arp~220 are inconsistent with the data in all twelve
objects studied.

\item Trends in the flux density ratios 350$\mu$m/24$\mu$m and
1200$\mu$m/24$\mu$m with the $R$-[24] color ($F_{\rm 24\mu m}/F_{\rm 0.7\mu
m}$) show that DOGs in this sample have elevated 24$\mu$m emission relative to
SMGs, most likely due to an obscured AGN.

\item The non-detections at 1mm imply $T_{\rm dust}$ greater than 35-60~K
for two objects.  

\item If the dust properties of the two DOGs observed at 1mm apply generally
    to the 24$\mu$m bright DOGs, then we estimate dust masses for these sources
    of $1.6-6.1\times10^{8} \: M_\sun$.  Lower $T_{\rm dust}$ would
    imply higher dust masses and vice versa.

\item In comparison to other $z \approx 2$ ULIRGs, DOGs have warmer dust
    temperatures that imply higher IR luminosities and lower dust masses.  This
    may be an indication that AGN growth has heated the ambient ISM in these
    sources.

\item Our stellar mass estimates provide weak evidence indicating that the
    24$\mu$m-bright DOGs may have converted more gas into stars than SMGs or
    other {\it Spitzer}-selected sources, consistent with them representing a
    subsequent phase of evolution.  An important caveat to this conclusion is
    that we have assumed DOGs and SMGs share the same gas mass to dust mass
    ratio.  Testing this assumption will require new data and will be an
    important goal of future work.

\end{enumerate}

This work is based in part on observations made with the {\it Spitzer Space
Telescope}, which is operated by the Jet Propulsion Laboratory, California
Institute of Technology under NASA contract 1407.  {\it Spitzer}/MIPS
guaranteed time observing was used to image the Bo\"otes field at 24$\mu$m and
is critical for the selection of DOGs.  We thank the SDWFS team (particularly
Daniel Stern and Matt Ashby) for making the IRAC source catalogs publicly
available.  Data from the original IRAC shallow survey were used for intial
stellar mass estimates.  We thank the anonymous referee for a thorough review
of the manuscript that helped improve the paper.

We are grateful to the expert assistance of the staff of Kitt Peak National
Observatory where the Bo\"{o}tes field observations of the NDWFS were obtained.
The authors thank NOAO for supporting the NOAO Deep Wide-Field Survey. In
particular, we thank Jenna Claver, Lindsey Davis, Alyson Ford, Emma Hogan, Tod
Lauer, Lissa Miller, Erin Ryan, Glenn Tiede and Frank Valdes for their able
assistance with the NDWFS data.  We also thank the staff of the W.~M.~Keck
Observatory, where some of the galaxy redshifts were obtained.

RSB gratefully acknowledges financial assistance from HST grant GO10890,
without which this research would not have been possible.  Support for Program
number HST-GO10890 was provided by NASA through a grant from the Space
Telescope Science Institute, which  is operated by the Association of
Universities for Research in  Astronomy, Incorporated, under NASA contract
NAS5-26555.  The research activities of AD are supported by NOAO, which is
operated by the Association of Universities for Research in Astronomy (AURA)
under a cooperative agreement with the National Science Foundation.  Support
for E. Le Floc'h was provided by NASA through the Spitzer Space Telescope
Fellowship Program.

Facilities used: {\it Spitzer}, CSO, and CARMA.  This research made use of CSO
(SHARC-II) and CARMA data.  Support for CARMA construction was derived from the
states of California, Illinois, and Maryland, the Gordon and Betty Moore
Foundation, the Kenneth T.  and Eileen L. Norris Foundation, the Associates of
the California Institute of Technology, and the National Science Foundation.
Ongoing CARMA development and operations are supported by the National Science
Foundation under a cooperative agreement, and by the CARMA partner
universities.

\end{document}